\documentclass[twoside,twocolumn,9pt]{article}
\usepackage{extsizes}
\usepackage[super,sort&compress,comma]{natbib} 
\usepackage[version=3]{mhchem}
\usepackage[left=1.5cm, right=1.5cm, top=1.785cm, bottom=2.0cm]{geometry}
\usepackage{balance}
\usepackage{times,mathptmx}
\usepackage{sectsty}
\usepackage{lastpage}
\usepackage[format=plain,justification=justified,singlelinecheck=false,font={stretch=1.125,small,sf},labelfont=bf,labelsep=space]{caption}
\usepackage{fnpos}
\usepackage{fancyhdr}
\usepackage[english]{babel}
\addto{\captionsenglish}{%
	
}
\usepackage{array}
\usepackage{droidsans}
\usepackage{charter}
\usepackage[T1]{fontenc}
\usepackage[usenames,dvipsnames]{xcolor}
\usepackage{setspace}
\usepackage[compact]{titlesec}
\usepackage{epstopdf}
\definecolor{cream}{RGB}{222,217,201}
\definecolor{darkblue}{RGB}{0, 0, 139}

\usepackage[utf8]{inputenc}
\usepackage{ucs}
\usepackage{placeins}
\usepackage{ulem}
\usepackage{amsmath}
\usepackage{amsfonts}
\usepackage{amssymb}
\usepackage{graphicx}
\usepackage{dcolumn}
\usepackage{bm}
\usepackage{latexsym}
\usepackage{hyperref}
\usepackage{color}
\usepackage{epsfig}
\usepackage{float}
\usepackage{wasysym}
\usepackage{fdsymbol}
\usepackage{color}
\usepackage{graphicx}
\usepackage{epsfig}
\usepackage{latexsym}
\usepackage{booktabs}
\usepackage{multirow}
\usepackage{fancyhdr}
\usepackage{wrapfig}
\usepackage{enumerate}
\usepackage{cancel}
\usepackage{mathrsfs}
\usepackage{mdwlist}
\usepackage{tikz}
\usepackage{amsmath}
\usepackage{amsfonts}
\usepackage{amssymb}
\usepackage{varwidth}

\begin{document}
	
	\pagestyle{fancy}
	\thispagestyle{plain}
	\fancypagestyle{plain}{
		
		\fancyhead[C]{\includegraphics[width=18.5cm]{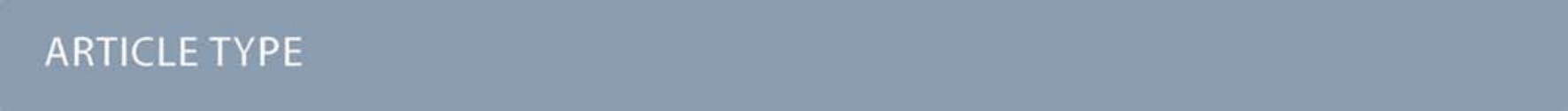}}
		\fancyhead[L]{\hspace{0cm}\vspace{1.5cm}\includegraphics[height=30pt]{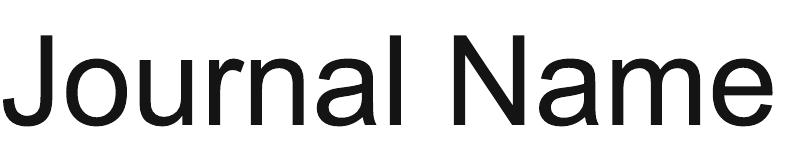}}
		\fancyhead[R]{\hspace{0cm}\vspace{1.7cm}\includegraphics[height=55pt]{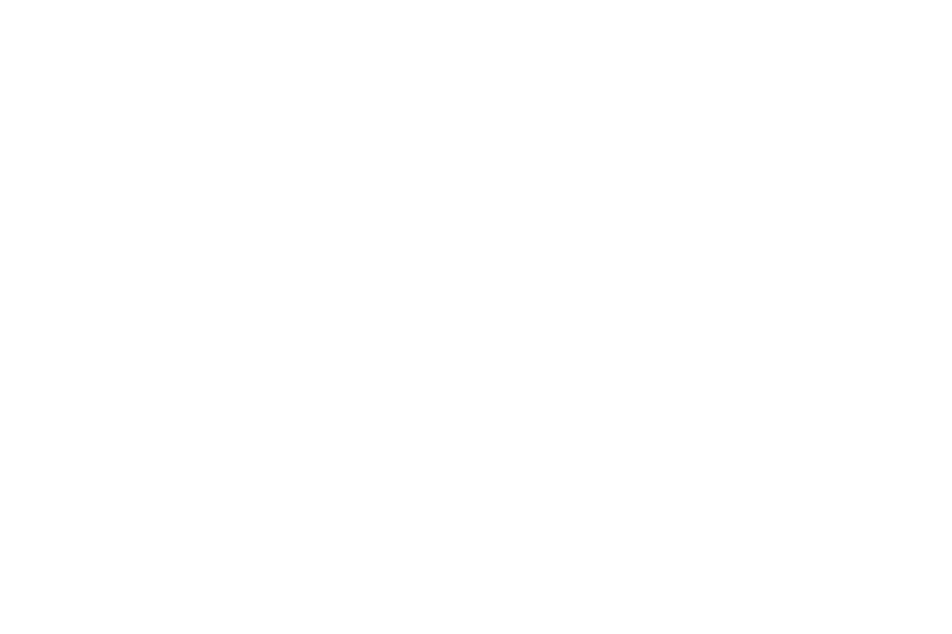}}
		\renewcommand{\headrulewidth}{0pt}
	}
	
	\makeFNbottom
	\makeatletter
	\renewcommand\LARGE{\@setfontsize\LARGE{15pt}{17}}
	\renewcommand\Large{\@setfontsize\Large{12pt}{14}}
	\renewcommand\large{\@setfontsize\large{10pt}{12}}
	\renewcommand\footnotesize{\@setfontsize\footnotesize{7pt}{10}}
	\makeatother
	
	\renewcommand{\thefootnote}{\fnsymbol{footnote}}
	\renewcommand\footnoterule{\vspace*{1pt}%
		\color{cream}\hrule width 3.5in height 0.4pt \color{black}\vspace*{5pt}} 
	\setcounter{secnumdepth}{5}
	
	\makeatletter 
	\renewcommand\@biblabel[1]{#1}            
	\renewcommand\@makefntext[1]%
	{\noindent\makebox[0pt][r]{\@thefnmark\,}#1}
	\makeatother 
	\renewcommand{\figurename}{\small{Fig.}~}
	\sectionfont{\sffamily\Large}
	\subsectionfont{\normalsize}
	\subsubsectionfont{\bf}
	\setstretch{1.125} 
	\setlength{\skip\footins}{0.8cm}
	\setlength{\footnotesep}{0.25cm}
	\setlength{\jot}{10pt}
	\titlespacing*{\section}{0pt}{4pt}{4pt}
	\titlespacing*{\subsection}{0pt}{15pt}{1pt}
	
	\fancyfoot{}
	\fancyfoot[LO,RE]{\vspace{-7.1pt}\includegraphics[height=9pt]{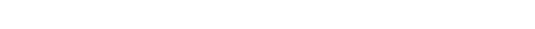}}
	\fancyfoot[CO]{\vspace{-7.1pt}\hspace{13.2cm}\includegraphics{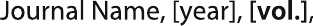}}
	\fancyfoot[CE]{\vspace{-7.2pt}\hspace{-14.2cm}\includegraphics{head_foot/RF}}
	\fancyfoot[RO]{\footnotesize{\sffamily{1--\pageref{LastPage} ~\textbar  \hspace{2pt}\thepage}}}
	\fancyfoot[LE]{\footnotesize{\sffamily{\thepage~\textbar\hspace{3.45cm} 1--\pageref{LastPage}}}}
	\fancyhead{}
	\renewcommand{\headrulewidth}{0pt} 
	\renewcommand{\footrulewidth}{0pt}
	\setlength{\arrayrulewidth}{1pt}
	\setlength{\columnsep}{6.5mm}
	\setlength\bibsep{1pt}
	
	\makeatletter 
	\newlength{\figrulesep} 
	\setlength{\figrulesep}{0.5\textfloatsep} 
	
	\newcommand{\topfigrule}{\vspace*{-1pt}%
		\noindent{\color{cream}\rule[-\figrulesep]{\columnwidth}{1.5pt}} }
	
	\newcommand{\botfigrule}{\vspace*{-2pt}%
		\noindent{\color{cream}\rule[\figrulesep]{\columnwidth}{1.5pt}} }
	
	\newcommand{\dblfigrule}{\vspace*{-1pt}%
		\noindent{\color{cream}\rule[-\figrulesep]{\textwidth}{1.5pt}} }
	
	\makeatother
	
	\twocolumn[
	\begin{@twocolumnfalse}
		\vspace{3cm}
		\sffamily
		\begin{tabular}{m{4.5cm} p{13.5cm} }
			
			\includegraphics{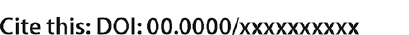} & \noindent\LARGE{\textbf{Phases and excitations of active rod-bead mixtures: simulations and experiments$^\dag$}} \\
			\vspace{0.3cm} & \vspace{0.3cm} \\
			
			& \noindent\large{Harsh Soni,\textit{$^{a,b}$} Nitin Kumar,\textit{$^{a,c}$}  Jyothishraj Nambisan,\textit{$^{a,d}$} Rahul Kumar Gupta,\textit{$^{b}$} A.K.
				Sood,\textit{$^{a}$}  and Sriram Ramaswamy\textit{$^{a,b}$}} \\
		
			\includegraphics{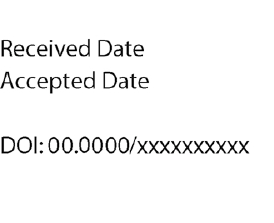} & \noindent\normalsize{
				We present a large-scale numerical study, supplemented by experimental observations, of a quasi-two-dimensional active system of polar rods and spherical beads confined between two horizontal plates and energised by vertical vibration. 				
				  For low rod concentrations $\Phi_r$ we observe a direct phase transition, as bead concentration $\Phi_b$ is increased, from the isotropic phase to a homogeneous flock. For $\Phi_r$ above a threshold value, an ordered band dense in both rods and beads occurs between the disordered phase and the homogeneous flock, in both experiments and simulations. Within the size ranges accessible we observe only a single band, whose width increases with $\Phi_r$. Deep in the ordered state, we observe broken-symmetry ``sound'' modes and giant number fluctuations. The direction-dependent sound speeds and the scaling of fluctuations are consistent with the predictions of field theories of flocking, but sound damping rates show departures from such theories. At very high densities we see phase separation into rod-rich and bead-rich regions, both of which move coherently. 
			} \\
			
		\end{tabular}
		
	\end{@twocolumnfalse} \vspace{0.6cm}
	
	]
	
	\renewcommand*\rmdefault{bch}\normalfont\upshape
	\rmfamily
	\section*{}
	\vspace{-1cm}

	\footnotetext{\textit{$^{a}$~Department of Physics, Indian Institute of Science, Bangalore
			560 012, India}}
	\footnotetext{\textit{$^{b}$~TIFR Centre for Interdisciplinary Sciences, Tata
			Institute of Fundamental Research, Hyderabad 500 107, India}}
	\footnotetext{\textit{$^{c}$~now at Department of Physics, Indian Institute of Technology Bombay, Powai, Mumbai 400 076, India}}
	\footnotetext{\textit{$^{d}$~now at School of Physics, Georgia Institute of Technology, 770 State Street NW, Atlanta, GA, 30332-0430, USA}}
	\footnotetext{\dag~Electronic Supplementary Information (ESI) available: [details of any supplementary information available should be included here]. See DOI: 00.0000/00000000.}
	

\def\gdot{\dot{\gamma}}
\def\vpl{v_\parallel}
\def\vperp{v_\perp}
\def\deg{^{\circ}}
\def\parpl{\partial_\parallel}
\def\parperp{\partial_\perp}
\def\otens{\mbox{\boldmath $\Omega$\unboldmath}}
\def\sigbold{\mbox{\boldmath $\sigma$\unboldmath}}
\def\Pibold{\mbox{\boldmath $\Pi$\unboldmath}}
\def\sigtens{\mbox{\boldmath $\sigma$\unboldmath}}
\def\sigac{\mbox{\boldmath $\sigma$\unboldmath$^a$}}
\def\signoise{\mbox{\boldmath $\sigma$\unboldmath$^n$}}
\def\lsim{\:\raisebox{-0.5ex}{$\stackrel{\textstyle<}{\sim}$}\:}
\def\gsim{\:\raisebox{-0.5ex}{$\stackrel{\textstyle>}{\sim}$}\:}
\def\3dots{\:\raisebox{-0.5ex}{$\stackrel{\textstyle.}{:}$}\:}
\def\beq{\begin{equation}}
\def\eeq{\end{equation}}
\def\bea{\begin{eqnarray}}
\def\eea{\end{eqnarray}}

\newcommand{\bsf}[1]{\textsf{\textbf{#1}}}
\newcommand{\hrs}[1]{\textcolor{black}{#1}}
\newcommand{\doubt}[1]{\textcolor{cyan}{#1}}
\newcommand{\SR}[1]{\textcolor{black}{#1}}
\newcommand{\remarkSR}[1]{\texttt{\textcolor{magenta}{#1}}}

\section{Introduction}
Active matter is the focus of intense current interest for its dramatic mechanical and statistical properties~\cite{RevModPhys.85.1143,srar,Toner2005170,BechingerRMP2016} such as giant number fluctuations~\cite{sr_epl_2003,Narayan06072007,vijay_thesis,tonertu_pre_1998}, wave propagation without conventional inertia~\cite{soundwave_prl_1997,tonertu_pre_1998,bartolo2018,toner2018}, broken continuous symmetry in two dimensions, band formation~\cite{chate_epjb_2008,PhysRevLett.110.208001,chate_bands_2014}, sustained spontaneous oscillations, instability of simple liquid-crystalline order in bulk fluid~\cite{sriram_aditi_2002}, motile topological defects~\cite{Narayan06072007,vijay_thesis,dogic2012,Marchetti2013PRL,Marchetti2014,SriramMarchettiBowick2018PRL,Doostmohammadi2018NatCom,NKumarNematics}, Motility-Induced Phase Separation \SR{and its generalisations~\cite{mips2015review,cates2019active,NKumarTrapping,redner2013structure}} and viscosity reduction through internally generated stresses~\cite{prl2015bacteriaViscosity,prl2009bacteriaViscosity,pre2011bacteriaViscosity}. The origin of these properties is the energy supply at the microscopic level, directly to the constituent particles, unlike in conventional nonequilibrium systems such as sheared fluids, which are powered through their periphery.
The energy transduction responsible for the active character of a particle could be wholly internal, as in living organisms~\cite{bray2000,alberts1994molecular,julicher2018hydrodynamic}, or it could take place in the region of contact of the particle with its surroundings, as with self-phoretic colloids in a fluid~\cite{PhysRevE.89.062316}, Quincke rollers~\cite{Bricard2013} or vibrated grains~\cite{Narayan06072007,vijay_jstat_theor_exp_2006_smectiv_nematic,kumar2014flocking}. Active particles are commonly elongated and can therefore form orientationally ordered states, of which the simplest are nematic and polar uniaxial liquid crystals. Diverse mechanisms underlie the ordering of active particles: in living active systems like bird flocks~\cite{cavagna2014bird}, fish schools and herds the process is behavioural and based on mutual sensing. In active granular systems, \SR{which exemplify dry active matter \cite{Chate2020Review},} steric and collisional effects give rise to alignment~\cite{Narayan06072007,polar_disk_long_soft_matter}. In colloidal rollers a restricted version of the hydrodynamic interaction is responsible~\cite{Bricard2013}.

Considerable progress has been made towards understanding active polar systems since the discovery of a flocking phase transition in an agent-based model~\cite{vicsek1995novel}, including field-theoretic arguments towards the existence of long-range order in 2D flocks \cite{PhysRevLett.75.4326,tonertu_pre_1998,Toner2005170,Toner_Malthus,toner2018}, and predictions of highly anisotropic sound waves -- through the interplay of the concentration and broken-symmetry fields --
and anomalously large number fluctuations in the ordered phase~\cite{soundwave_prl_1997,tonertu_pre_1998,Toner2005170}. Recently, Geyer~\textit{et al.}~\cite{bartolo2018} observed sound waves in the suspension of active colloidal rollers. 
Bertin \textit{et al.}~\cite{bertin_pre_2006}, in a Boltzmann-equation construction of the Toner-Tu equations, discovered that the dependence of the local ordering tendency on the local density inevitably led to a linear instability of the ordered phase, with wavevector parallel to the ordering direction, just past the mean-field flocking transition. Indeed, a banded phase is widely observed to intervene between the isotropic and the uniform ordered phases in agent-based numerical simulations~\cite{chate_epjb_2008,PhysRevLett.110.208001,chate_bands_2014} and in experiments on rolling-colloid flocks~\cite{Bricard2013}, and arises as well in a variety of theoretical models~\cite{active_ising_prl_2013,toner_band_2014_prl,hydro_theory_band_prl_2014_Tailleur}.

Here we study a two-dimensional active polar monolayer consisting of tapered rods adrift in a sea of spherical beads. 
The energy input to the particles is provided by a vertically vibrated supporting surface, and the rods, by virtue of their shape, transduce this vibration into directed movement in the plane. The spherical beads mediate an aligning interaction between the rods, and otherwise behave like passive particles that move if pushed or dragged by the polar rods. In earlier work on this system~\cite{kumar2014flocking,harsh_thesis,nitin_thesis},  
we discovered a nonequilibrium phase transition from the isotropic state to an ordered, coherently moving flock,  
which took place when the concentration of the spherical beads exceeded a critical value which decreased with increasing concentration of rods. Our experimental results were supported by a hydrodynamic theory and numerical simulations incorporating the detailed Newtonian mechanics of the particles and boundaries, including vibration, inelasticity and static friction. In this paper, we offer a a detailed exploration of the phase diagram, mode structure and spatiotemporal correlations of this system, primarily in simulations but supported by key experimental findings. There are of course many parameters one could consider varying, such as the concentrations of rods and beads, the rotational diffusivity of the rods, coefficients of friction between the particles and the substrate, several of which feed into the speed of the rods, and thus mainly change the effective clock speed of the dynamics. We restrict our studies to the dependence on rod area fraction $\Phi_r$ and the bead area fraction $\Phi_b$. Our numerical studies all employ periodic boundary conditions (PBC) in the horizontal plane, thus eliminating the role of the lateral boundaries of the sample.

Here is a summary of our main results. \SR{(i) Large-scale simulation studies over a range of $\Phi_r$ and $\Phi_b$, with periodic boundary conditions, reveal that a banded state intervenes between the isotropic state and the homogeneous ordered phase, for $\Phi_r$ above a threshold value $\Phi^c_r$, as $\Phi_b$ is increased. We see a single coherently moving band rich in both rods and beads, not a periodic array. Detailed studies of band morphology can be found in the body of the paper. Below $\Phi^c_r$, the system appears, within our resolution, to undergo a phase transition from a disordered to an ordered state directly, as discussed in~\cite{kumar2014flocking}. 
(ii) We present experimental evidence for a banded flock, to our knowledge for the first time in dry granular matter. (iii) In the homogeneous ordered state our numerical studies reveal giant number fluctuations and a spectrum of propagating modes. The observed wavevector dependence of the damping rates of the modes differs from the predictions of the Toner-Tu theory, which does however capture other broad features such as the scaling of the number fluctuations, and direction-dependent wavespeeds. (iv) At higher values of total concentration of the rods and the beads we observe phase segregation into bead-rich and aligned rod-rich regions, both moving coherently. 
}
The remainder of the paper is organised as follows: in section~\ref{methods}, we discuss our numerical and experimental methods. In section~\ref{results} we present our detailed results. We summarise and suggest future directions in the last section~\ref{summary}.

\section{Methods}\label{methods}  

\subsection{Experiment}
Our experimental cell is a shallow circular geometry, made of
hardened aluminium alloy. The particles are confined to two dimensions using a glass lid, which is fixed on the external perimeter of the circle at a height of $w=$1.2 mm above
the base. We use the ``flower'' geometry \cite{PhysRevLett.105.098001,kumar2014flocking} to prevent clustering of particles on the cell boundary. The cell is mounted on a permanent magnet shaker (LDS 406/8) and
is shaken at a fixed frequency $f=200Hz$ and amplitude $a_{0}$. The amplitude of the resulting sinusoidal acceleration $\Gamma \equiv a_{0}(2 \pi
f)^{2}/g$, measured by an accelerometer (PCB Piezoelectronics 352B02), is chosen to be 7.0 in the units of the gravitational acceleration of the earth $g$. Our ``self-propelled'' polar particle, which we call  a ``rod'' henceforth,  is a brass rod, \hrs{$\ell=$4.5 mm long} and 1.1 mm in diameter at its thick end as shown in Fig. 1(a). The tilt of the rod with respect to the horizontal transduces the energy of vertical vibration into in-plane propulsion and its geometrical polarity, that is fore-aft asymmetry, ensures that this propulsion is biased towards one end of the rod, specifically, the narrow end. The other particles in our experiment are spherical beads of aluminium, 0.8 mm in diameter, which do not show any in-plane dynamics when vibrated vertically. We created an annular geometry by inserting a circular disk of 5 cm diameter in the middle of the experimental cell. We expected this geometry to stabilize flocking along the azimuthal direction and therefore to favour bands.

\begin{figure}[htp]
	\centering \includegraphics[width=0.5\textwidth]{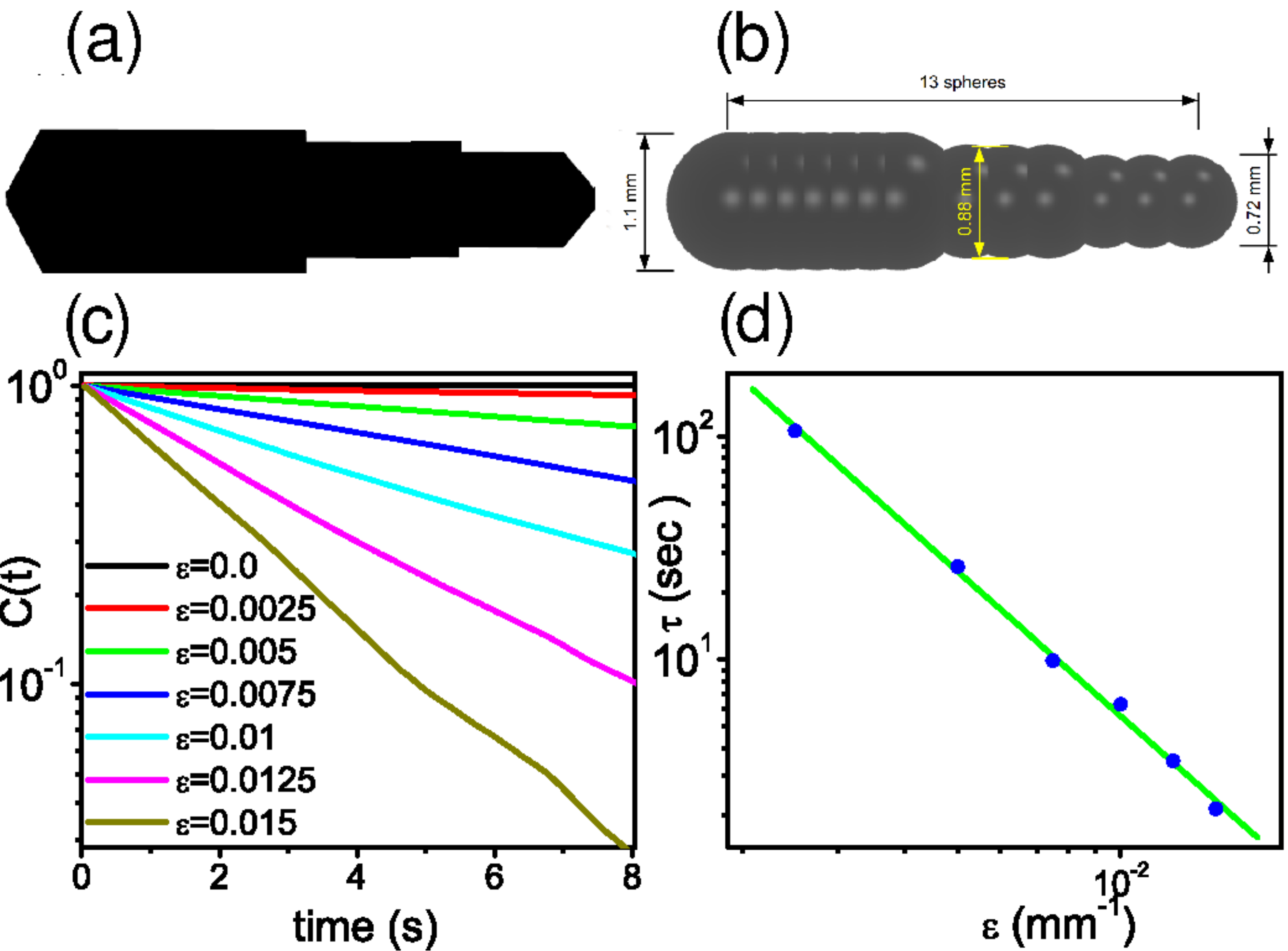} 
	\caption{(\textbf{a}) A schematic diagram of the experimental rod. (\textbf{b}) The simulated rod which is made of 13 spheres overlapping with each other.\textbf{(c)} The orientation autocorrelation function $C(t)$ of a single rod in simulations vs $t$ for different values of $\varepsilon$. Clearly, $C(t)\sim\exp(-t/\tau)$. \textbf{(d)} Relaxation time $\tau$ as the function of $\varepsilon$, showing as $ \varepsilon^{-2}$ dependence.}
	\label{method}
\end{figure}  

\begin{figure*}[t]
	\begin{center} 
		\includegraphics[width=1.0\textwidth]{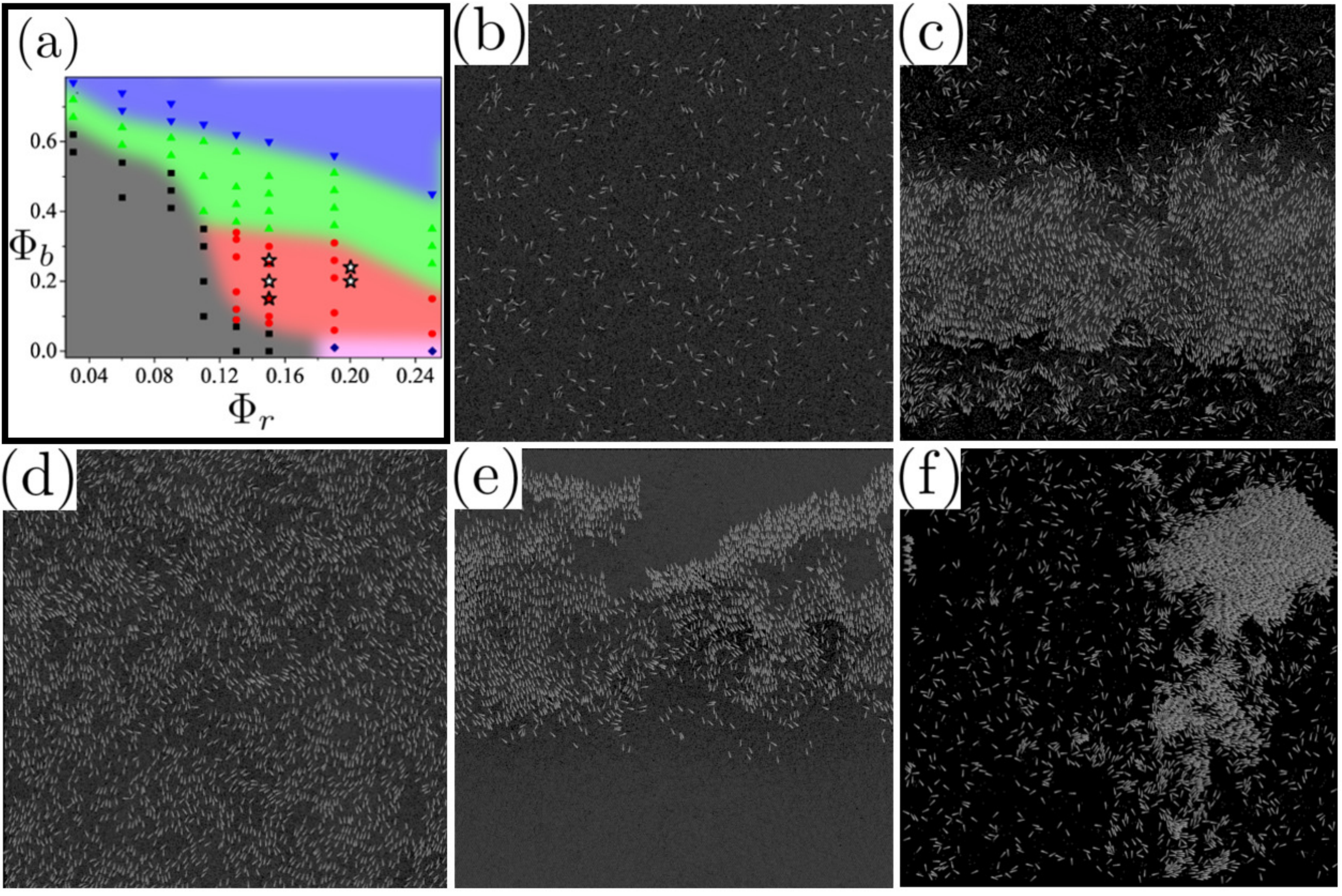} 
		\caption{\textbf{(a)} A phase diagram in the $\Phi_r-\Phi_b$ plane for the system size \hrs{$L = 56\ell$} (around 228 in diameters of the rod). \SR{Here we find bead-rod phase separation ($\textcolor{blue}{\blacktriangledown}$), ordered flock ($\textcolor{green}{\blacktriangle}$), disordered phase($\blacksquare$), bands ($\textcolor{red}{\CIRCLE}$) and swarms ($\textcolor{darkblue}{\medblackdiamond}$). The stars ($\bm{\largewhitestar}$) depict the location of the band phase in experiments with the annular geometry (see Fig.~\ref{expxpband}). \textbf{(b)} Isotropic state  at  $\Phi_r=0.03$ and $\Phi_b=0.57$. \textbf{(c)} A band surrounded by the isotropic state at $\Phi_r=0.19$ and $\Phi_b=0.21$. \textbf{(d)} A homogeneous ordered state at $\Phi_r=0.19$ and $\Phi_b=0.51$. \textbf{(e)} Bead-rod phase separation at very high density ($\Phi_r=0.15$ and $\Phi_b=0.60$).} 
			\textbf{(f)} Aligned swarm at $\Phi_r=0.19$ and a very low $\Phi_b=0.01$.
		}
		\label{phasediagram}
	\end{center}
\end{figure*}

\subsection{Simulation}\label{simmethods}
Our numerical simulations are based on a mechanically faithful reproduction of the microscopic dynamics of each particle. We assume that all the particles and walls are perfectly rigid. Therefore, all interactions are instantaneous events. All the collisions are inelastic with prescribed restitution and Coulomb friction, and the gravitational acceleration of the earth is taken into account in our simulations. The vibrating base and lid are modelled as horizontal walls moving in the vertical direction with their $z$-coordinates changing with time $t$ as $a_0\cos 2 \pi f t+a_0 $ and $a_0\cos 2 \pi f t+a_0+w $ respectively. We do not use the event-driven method~\cite{poschel:compgrandyn2004book}, often preferred for granular systems at low density, but use instead a time-driven algorithm~\cite{tagkey2002589}. The latter is more appropriate for our dense system, where an event-driven approach would require a large number of computations to predict the time of the next particle-particle and particle-wall collision~\cite{harsh_thesis}. 
Since the rod in experiments has a complicated shape(see Fig. \ref{method}a), several calculations are required to detect the collisions between the rods. Therefore, in order to simplify the collision rules and speed up the collision detection process, we construct the rod as an array of overlapping spheres (see Fig. \ref{method}b).

The ballistic motion of the particles is governed by Newtonian rigid body dynamics. The Impulse-Based Rigid Body Collision
Model~\cite{stronge,9780511626432} is implemented to calculate post collision velocities for all the collisions. We write an MPI-based parallel code to simulate our system: the simulation box is divided into many equal-sized sub-boxes and the dynamics of the particles in different sub-boxes are dealt with by different computer cores. At each step, data for the particles at the boundaries of each sub-box are communicated to the neighbour sub-boxes to execute the collisions between the particles across the sub-boxes. We use VMD software~\cite{VMD} to construct all the simulation movies and snapshots. To achieve the best imitation of the single-particle dynamics of rods and the beads in experiments, we choose the following values of restitution and friction coefficients $\mu$ and $e$:
\begin{table}[h]
	\small
	\caption{\ Values of restitution and friction coefficients}
	\label{tbl:example}
	\begin{tabular*}{0.48\textwidth}{@{\extracolsep{\fill}}lll}
		\hline
	Collision & $\mu$ & $e$ \\
		\hline
		Particle-particle & 0.05 & 0.3  \\
		Rod-base(or lid) & 0.03 & 0.1  \\ 
		Rod-boundary & 0.01 & 0.3  \\ 
		Bead-wall(base,lid or boundary) & 0.01 & 0.3  \\ 
		\hline
	\end{tabular*}
\end{table}
\begin{figure*}[htp] 
	\centering \includegraphics[width=1.0\textwidth]{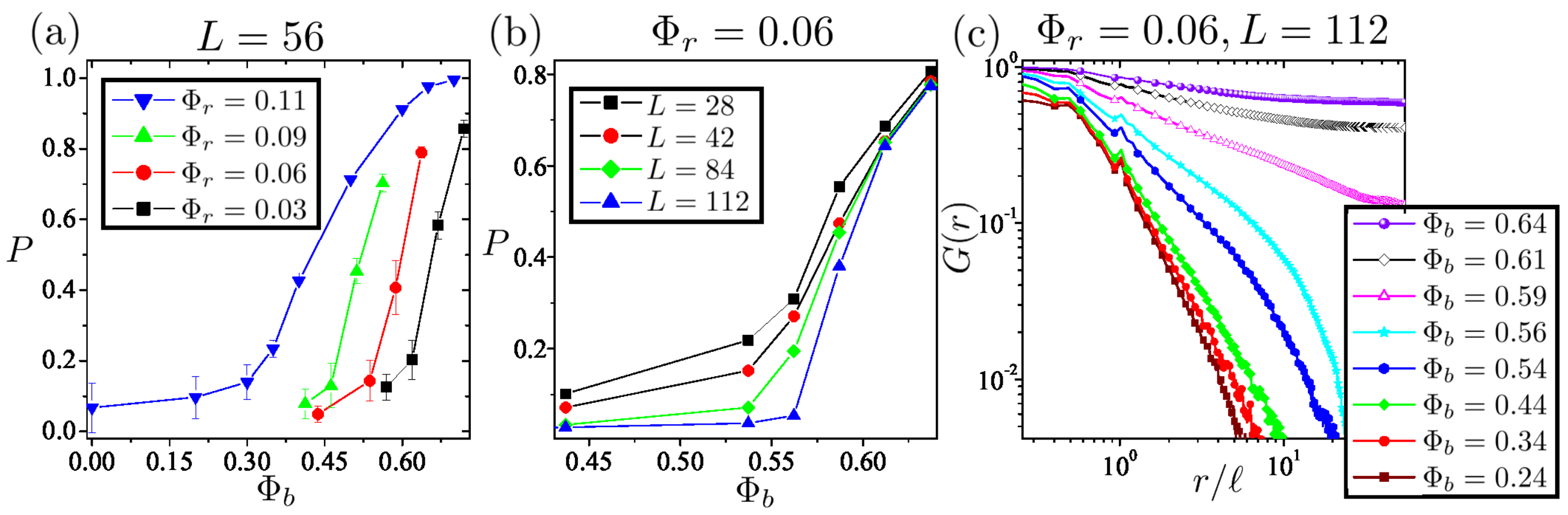} 
	\caption{ \textbf{(a)} Order parameter $P$ as the function of rod area fraction $\Phi_b$ for $\Phi_r=0.03,0.06,0.09$ and 0.11. \textbf{(b)} $P$ vs $\Phi_b$ for different system sizes at $\Phi_r=0.06$. \textbf{(c)} Behaviour of $G(r)$ as the function of $r$ for the largest system size ($L=112\ell$ or 457 rod diameters).}
	\label{figs}
\end{figure*}
We choose the size of the spheres such that the rod in simulations is a close mimic of the original rod in experiments: the tail of the rod is made of seven spheres of $1.1$ mm diameter, and the head and the middle parts each consist of three spheres of diameter $0.72$ mm and $0.88$ mm respectively (see Fig \ref{method}b). The beads are represented by spheres of diameter $0.8$ mm. The mass densities of the rods and beads are  8.7 gm/cm$^3$ and 2.7 gm/cm$^3$ corresponding to brass and aluminium respectively.  
The values of $w$, $f$ and $a_0$ are set to 1.2 mm, 200 Hz and 0.04 mm respectively, as in experiments. In experiments, imperfections in shape of the rods and the substrate roughness lead to the diffusive nature of the orientation of the rods. Numerical simulations lack such imperfections. Therefore, we supply the rods with noisy angular velocity $\omega_{z}=\varepsilon v_{rel} \eta$ in the $z$ direction  each time they collide with the base or the lid. Here, $ v_{rel}$ is the relative velocity at the contact point normal to the contact plane, $\varepsilon$ is a control
parameter and $\eta = \pm 1$ with equal probability. Fig \ref{method}c shows that the orientation autocorrelation function $C(t)$ of a single rod decays exponentially with time $t$, and the relaxation time $\tau$ is decreased with $\varepsilon$, going as  $\sim \varepsilon^{-2}$(see Fig.~\ref{method}d). The value of $\varepsilon$ is set to 0.015, as in~\cite{kumar2014flocking}.

\section{Results}~\label{results}
\hrs{We first present our findings from the numerical simulations and then discuss the bands seen in our experiments. In the rest of the paper, all lengths are scaled by the rod length $\ell$.}
\subsection{Simulations}~\label{simu}
\hrs{The simulations are performed with the periodic box of dimension $L=56\ell$ unless otherwise mentioned. The largest systems we studied had $L = 112 \ell$, with 7,200 rods and 300,000 beads.
We here discuss the observed phase diagram (see~\ref{phasedigram}), examine the ordering transition at small $\Phi_b$ (see~\ref{lowphi}), study the properties of bands (see~\ref{band_section}), 
and then present the sound wave spectrum (see~\ref{wave_section}) and large density fluctuations (see~\ref{largN}) in the highly ordered phase.}

\subsubsection{Phase diagram}~\label{phasedigram}
Fig. \ref{phasediagram}a, a phase diagram in the plane of  $\Phi_r$ and $\Phi_b$,  gives an overview of the behaviour of  the system in various regimes. We mainly observe four phases in our system: disordered, homogeneous ordered, banded, and phase-separated (see Fig~\ref{phasediagram}b-e). The disordered phase, which is found at small values of $\Phi_r$ and $\Phi_b$, is structureless at very low rod densities (see Fig.~\ref{phasediagram}b) but displays a few randomly moving highly dense and locally ordered swarms at higher values of $\Phi_r$ (see Fig.~\ref{disorder13_07} of Appendix), presumably incomplete MIPS~\cite{cates2015motility,separationfoot}.
For $\Phi_r > \Phi^c_r \simeq 0.13$ the band phase is present between the ordered and disordered state 
(see Fig.~\ref{phasediagram}c). In~\ref{band_section},  we present detailed observations on this phase. Ordered states are observed at higher values of $\Phi_b$ (see Fig.~\ref{phasediagram}d).
The velocity field of the bead medium plays an important role in achieving the ordered state~\cite{kumar2014flocking}, as can be seen strikingly through a careful choice of initial conditions. When we perform the simulation of a system which is initially at rest, with the rods in an aligned state, the rods immediately start moving with a constant speed but the beads take some time to pick up their steady state in-plane velocity. Therefore, initially the bead flow is not enough to keep the rods ordered, and the polar order parameter decreases because the rods start disordering. Since they are dragged and pushed by the rods, the beads acquire some speed after some time, and the polar order parameter again increases due to the ordering enhanced by the bead flow (see Fig.~\ref{rolebeads} of Appendix).
More interestingly, the rods can flock at ultra-low $\Phi_r$ (as low as $ \approx 0.03$) if $\Phi_b$ is high enough (see Fig. \ref{r03t75} of Appendix). At very high densities, phase segregation into bead-rich and (ordered) rod-rich regions is observed (see Fig. \ref{phasediagram}e and Supplementary Movie 1). This regime occurred in our earlier experiments \cite{kumar2014flocking}, where it was characterised as jamming because the motile rods were immobilised as they pushed up against the sample boundary. A detailed study of the system in this regime will be presented in a separate paper. At higher values of $\Phi_r$ ($\gtrsim 0.19$) the rods are condensed into large dense ordered swarms for $\Phi_b \ll 1$, through a MIPS-like mechanism (see Fig. \ref{phasediagram}f).
  
Finite-size effects are significant in our study. The observed value of $\Phi^c_r$ and the boundaries of the phases are influenced by system size $L$. We find that $\Phi^c_r$ decreases as $L$ is increased but we find no banded state at $\Phi_r=0.06$ even for the system size as big as $L= 112\ell$  (or around 457 rod diameters). \SR{We do not pursue the question of whether} $\Phi^c_r$ can reach zero for very large systems, as seen in agent based numerical simulations~\cite{chate_bands_2014}. 
 
\begin{figure*}[htb] 
	\centering \includegraphics[width=1.0\textwidth]{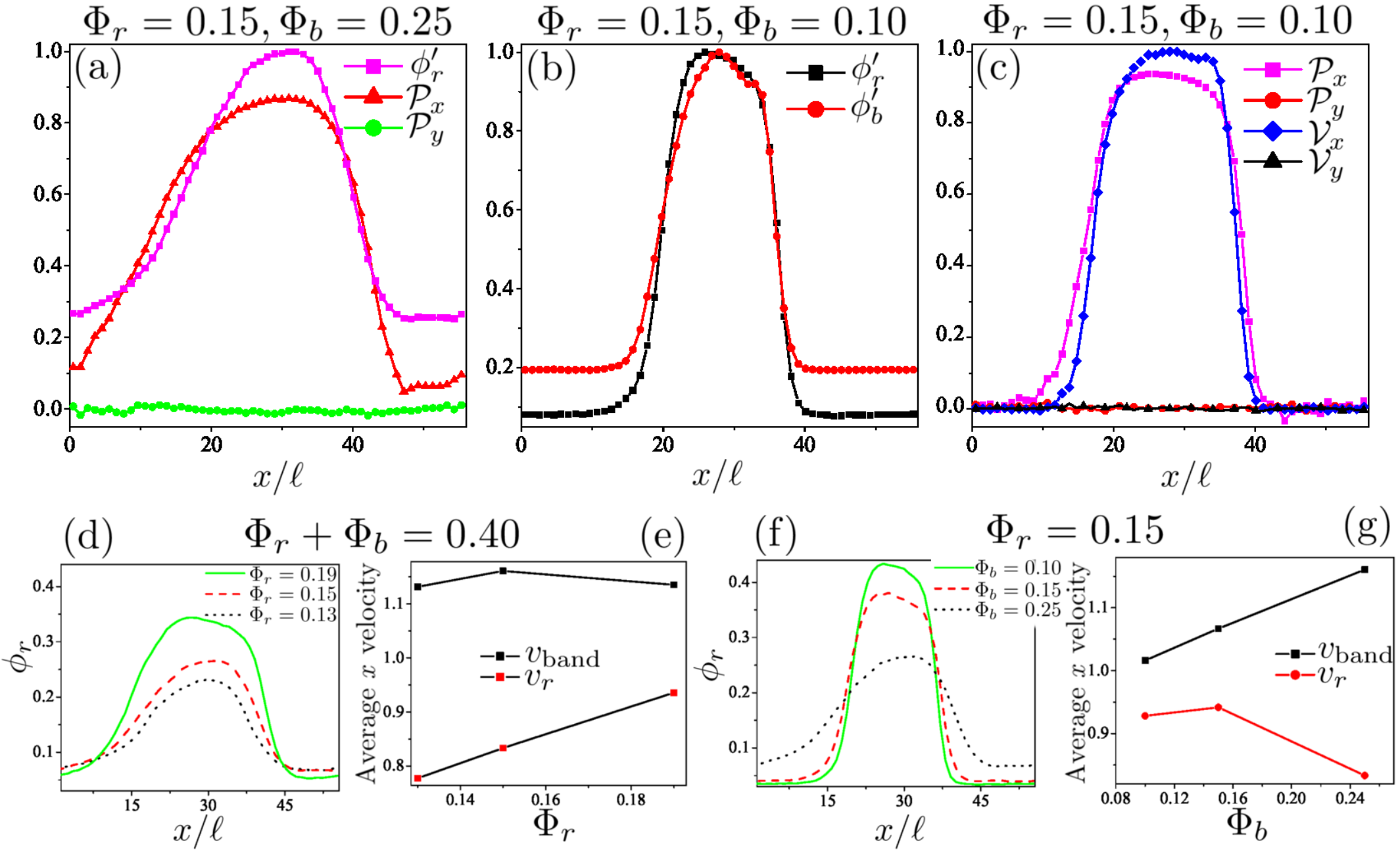} 
	\caption{
		Characteristics of the bands moving in the $x$ direction. \textbf{(a)} The scaled density profile $\phi'_r(x)\equiv\phi_r(x)/\textrm{max}[\phi_r(x)]$ and ordering profile $\bm{\mathcal{P}}(x)$ of the rods for the band at $\Phi_r=0.15$ and $\Phi_b=0.25$. \textbf{(b)} The scaled density profile for the rods and the beads  $\phi'_r(x)$ and  $\phi'_b(x)\equiv\phi_b(x)/\textrm{max}[\phi_b(x)]$  at $\Phi_r=0.15$ and $\Phi_b=0.10$. \textbf{(c)} Ordering profile of the rods $\bm{\mathcal{P}}(x)$ and the velocity profile of the beads $\bm{\mathcal{V}}(x)$ at the parameter values as in (b). At $\Phi_r+\Phi_b=0.40$, for different values of $\Phi_r$:  \textbf{(d)} The rod density profile $\phi_r(x)$ for different values of $\Phi_r$, and \textbf{(e)} The average velocity of the bands $v_{band}$ and average velocity of their constituent rods $v_{r}$ as the function of $\Phi_r$. At $\Phi_r=0.15$, for different values of $\Phi_b$: \textbf{(f)} The rod density profile $\phi_r(x)$ for different values of $\Phi_b$.  \textbf{(g)}  $v_{band}$ and $v_{r}$ as the function of $\Phi_b$. Here $L=56$.
	}
	\label{band1}
\end{figure*}

\subsubsection{Phase transition at low $\Phi_r$}~\label{lowphi}
In order to measure the ordering of the rods, we first define the order parameter of the rods as  $P\equiv|\langle\textbf{n}_i(t)\rangle|$, where $\textbf{n}_i(t)$ is the orientation unit vector of the $i$th rod and $\langle\rangle$ denotes the average over all rods and many configurations in the steady state.
Our simulations suggest a direct disorder-to-order phase transition without bands at low $\Phi_r(<\Phi^c_r)$, although we cannot strictly rule out bands on a much larger length scale. In Fig. \ref{figs}a we plot the order parameter $P$ as a function of  $\Phi_b$ for $\Phi_r=0.03,0.06,0.09$ and 0.11.: $P$ visibly increases with $\Phi_b$.  In Fig \ref{figs}b  $P$ is plotted as the function $\Phi_b$ for  $L=28$, 42, 84 and 112, at $\Phi_r=0.06$. The profile becomes sharper with increasing $L$ and  the graphs for different $L$ don't intersect with each other, this is consistent with a continuous phase transition, but our size range is limited to conclude. We now calculate the polar order parameter correlation function defined as $G(r)\equiv\langle\textbf{n}_i(t).\textbf{n}_j(t)\rangle_r$. Here, the averaging is performed our all the pairs of the rods separated by distance $r$. Fig. \ref{figs}c, which shows  $G(r)$ vs $r$ for different values of $\Phi_b$ at $\Phi_r=0.06$ and L = 112, illustrates that $G(r)$ decays to a nonzero constant at high $\Phi_b$ corresponding to the long range ordering, and vanishes for large $r$ at low $\Phi_b$ in disordered states ~\cite{kumar2014flocking}. Thus, our numerical experiments are consistent with long-range ordering in two dimensions as argued by Toner \textit{et al.}~\cite{PhysRevLett.75.4326,Toner_Malthus}.

\subsubsection{Properties of bands}\label{band_section}

At high rod densities $(\Phi_r \geq \Phi^c_r)$ a phase is seen between the order and the disordered states in which a single highly ordered and highly dense stripe of the rods and the beads extended over the length of the simulation box -\textit{band}- is observed to be  moving perpendicular to its own long axis, amidst a disordered background also consisting of of a bead-rod mixture (see Fig. \ref{phasediagram}e, and Supplementary Movie 2). In this system, up to $L=112$,  the segregation results in a single band unlike the periodically arranged many bands demonstrated by Vicsek particles \textit{et al.}~\cite{chate_epjb_2008,chate_bands_2014,chate_2008_pre_detailed}. The bands are generally  aligned along the sides of the simulation box but also could be in arbitrary direction  at large values of $\Phi_r$ (see Fig. \ref{diagonalbands} of Appendix). For convenience of analysis we study the regime in which the bands are parallel to  the length of the simulation box. Let $\textbf{R}^r_i(t)$ and  $\textbf{R}^b_i(t)$ be the positions  of $i$th rod and $i$th  bead respectively, and $\textbf{V}_i(t)$ the velocity of the $i$th bead, at time $t$. Respectively, the coarse grained number densities for the rods and beads are defined as
\begin{eqnarray}
\sigma(\textbf{r},t)=\frac{1}{l^2}\int_{\mbox{cell}} d^2{r} \sum_{i} \delta(\textbf{r}-\textbf{R}^r_i(t)),\\
\rho(\textbf{r},t)=\frac{1}{l^2}\int_{\mbox{cell}} d^2{r} \sum_{i} \delta(\textbf{r}-\textbf{R}^b_i(t)),
\end{eqnarray}
where $\sum$ stands for a sum over all the particles and the integration is taken over a square cell of length $l$ centred at position $\textbf{r}$. Similarly, the polar order parameter field for the rods and the velocity field  for the beads are given by 
\begin{eqnarray}
\textbf{p}(\textbf{r},t)=\frac{1}{l^2 \sigma(\textbf{r},t)} \int_{\mbox{cell}}  d^2{r} \sum_{i} \delta(\textbf{r}-\textbf{R}^r_i(t)) \textbf{n}_i(t),\\
\textbf{v}(\textbf{r},t)=\frac{1}{ l^2 \rho(\textbf{r},t)} \int_{\mbox{cell}}  d^2{r} \sum_{i} \delta(\textbf{r}-\textbf{R}^b_i(t)) \textbf{V}_i(t).
\end{eqnarray}
In order to quantify the density  and ordering profile of the rods we define
d $\phi_r(x)=\langle\sigma (\textbf{r},t)\rangle$ and $\bm{\mathcal{P}}(x)\equiv ({\mathcal{P}}_x(x),{\mathcal{P}}_y(x))=\langle \textbf{p} (\textbf{r},t)\rangle $, where we assume that the band is moving along the X axis and the angle bracket represents the average over the $Y$ direction and time in a frame moving with the band. Similarly, the density profile and the velocity profile of the beads are measured by $\phi_b(x)=\langle\rho (\textbf{r},t)\rangle$ and $\bm{\mathcal{V}}(x)\equiv ({\mathcal{V}}_x(x),{\mathcal{V}}_y(x))=\langle \textbf{v} (\textbf{r},t)\rangle /\textrm{max}[|\langle \textbf{v} (\textbf{r},t)\rangle|]$. We divide the simulation box into the cells of length one to calculate the value of these functions.
 \begin{figure} 
 	\centering \includegraphics[width=0.4\textwidth]{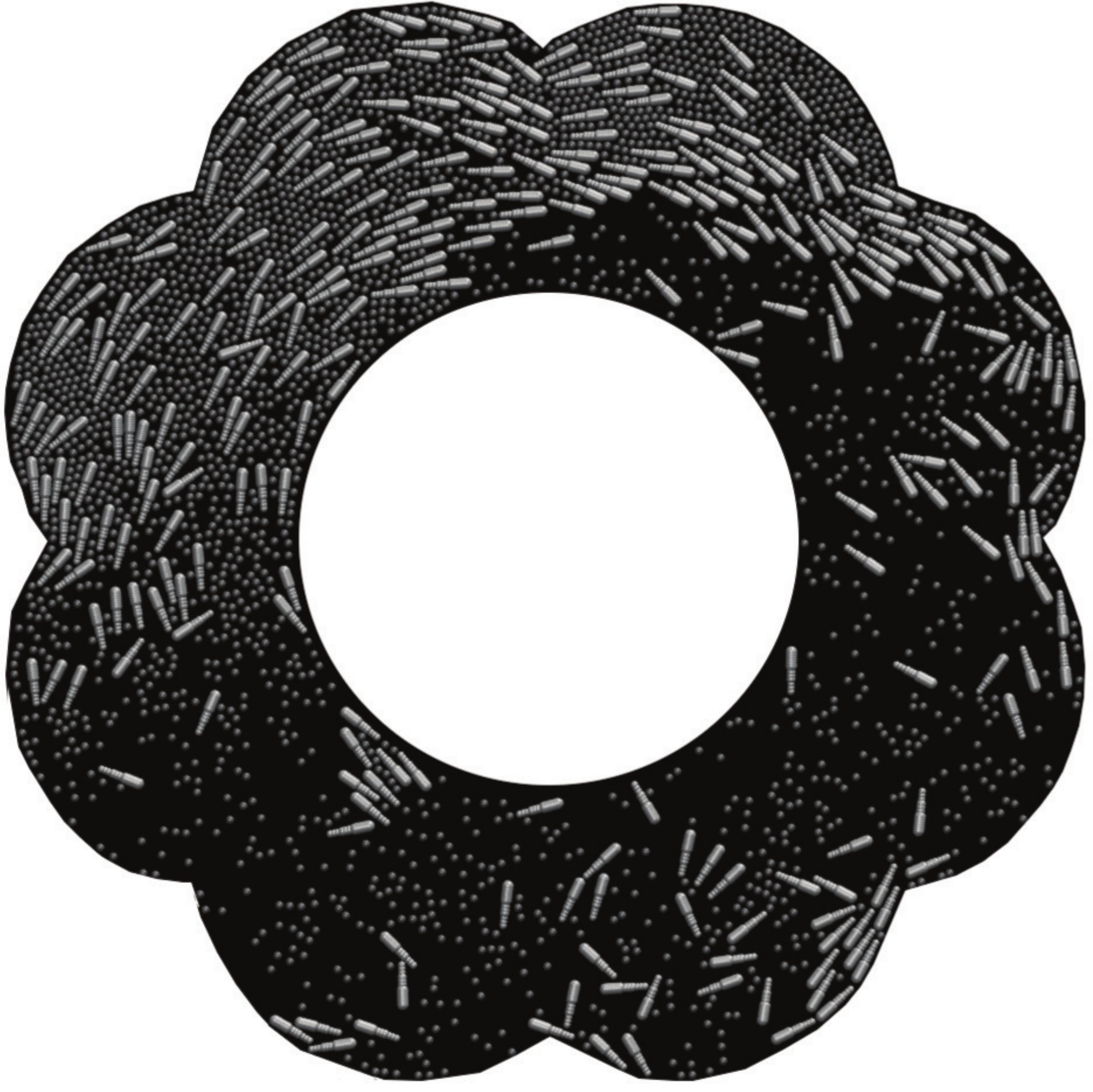} 
 	\caption{The banded structure seen in the simulations with the annular geometry having outer boundary same as that in our experiments. Here $\Phi_r=\Phi_b=0.20$.}
 	\label{simuexpband}
 \end{figure}
A graphical representation of the bands can be found in Fig. \ref{band1}a showing the typical density profile of the rods in a band moving in the $X$ direction:  the scaled density profile of the rods $\phi'_r(x)\equiv\phi_r(x)/\textrm{max}[\phi_r(x)]$  decays faster at the front than at the back \textit{ i.e.} the band is asymmetric with the front sharper than the back. The value of ${\mathcal{P}}_x(x)$ is close to 1 in the band region and fluctuates around 0 elsewhere, and ${\mathcal{P}}_y(x)$  remains close to zero everywhere in the simulation box, indicating that the rods in the band region are aligned  along the direction of the motion of the band and randomly oriented elsewhere. An interesting feature of these bands is that even the medium particles, the beads, form bands co-centred with the band of the rods (see Fig. \ref{band1}b), which is an aspect that cannot arise in the related colloidal rollers~\cite{Bricard2013}. In Fig. \ref{band1}c we plot $({\mathcal{V}}_x(x),{\mathcal{V}}_y(x))$ and   $({\mathcal{P}}_x(x),{\mathcal{P}}_y(x))$  as the function of $x$, for $\Phi_r=0.15$ and $\Phi_b=0.10$. Again in the band region the average velocity of beads is high and along the direction of the band and vanishingly small elsewhere. Also the profile of $\bm{\mathcal{V}}(x)$ is quite similar to of $\bm{\mathcal{P}}(x)$. We further explore the effect of the $\Phi_r$ and total area fraction $\Phi_t=\Phi_r+\Phi_b$ on the bands. The band  becomes wider and denser with increasing $\Phi_r$ at fixed $\Phi_t$ (see Fig.~\ref{band1}d). The condensation of the particles in a single band has also been  observed in the active Ising systems~\cite{active_ising_prl_2013,chate_bands_2014},  but,  in contrast, more than one band were observed in Vicsek systems \cite{chate_bands_2014} with the number of bands increasing with system size and particle density. The velocity of the bands $v_{band}$ doesn't change significantly with $\Phi_r$ but the average velocity of the rods lying  in the band region $v_r$ increases, probably because of the suppression of the transverse fluctuations of the orientation of the rods due to the  increasing density in the band region (see Fig.~\ref{band1}e). The band moves faster than the rods occupying the band because of the significant velocity gradient at its boundaries. For a given value of  $\Phi_r$, the  band widens with increasing $\Phi_b$, suggesting a tendency to dissolve the band into a homogeneous ordered state at high enough $\Phi_b$  (see Fig.~\ref{band1}f). Correspondingly,  $v_r$ goes down with $\Phi_b$ due to enhancement in the transverse fluctuations and $v_{band}$ rises due to decreasing  the density gradient across the band boundaries (see Fig.~\ref{band1}g). A shoulder-like trend is found in $G(r)$ vs $r$ as a result of the rectangular shape of the bands, indicating the typical width of the band (see Fig.~\ref{bandallgr} of Appendix). The nature of the phase transition at the boundaries of the segregated regime remains unclear.

Bands in our system do not appear to arise through the instability proposed in \cite{bertin_pre_2006}. Supplementary Movie 2 suggests a different mechanism. Initially small swarms with internal alignment are observed, moving randomly in the isotropic background, with both rods and beads joining and leaving them at their boundary. When the swarms come alongside one another, they unite through lateral exchange of particles to form a band.

\hrs{In order to explore the possibility of the bands in experiments, we execute the simulations with the annular geometry with outer boundary same as in our experiments~\cite{kumar2014flocking}. Fig.~\ref{simuexpband} and Supplementary Movie 3 show that the bands should be seen in experiments as well. We will present our experimental result on the bands in subsection~\ref{exp}. 
}

\subsubsection{Sound waves}\label{wave_section}
Fig. \ref{phasediagram}f and Supplementary Movie 4 are evidence for the propagating waves predicted by Toner \textit{et al.}~\cite{tonertu_pre_1998} in the uniform ordered state. In order to measure these waves we calculate the dynamic and static orientation structure factors. Let  $\textbf{n}^{\perp}_{i}(t)$ be the component of the orientation of $i$th rod normal to the direction of the flock. 
The field for the normal orientation fluctuations is then given by
\begin{eqnarray}
\delta \textbf{p}_{\perp}(\textbf{r}, t)=\dfrac{\sum_i \textbf{n}^{\perp}_{i}(t) \delta(\textbf{R}^r_i(t)-\textbf{r})}{\sum_i  \delta(\textbf{R}^r_i(t)-\textbf{r})},
\end{eqnarray}
\begin{figure}[H]
	\centering \includegraphics[width=0.5\textwidth]{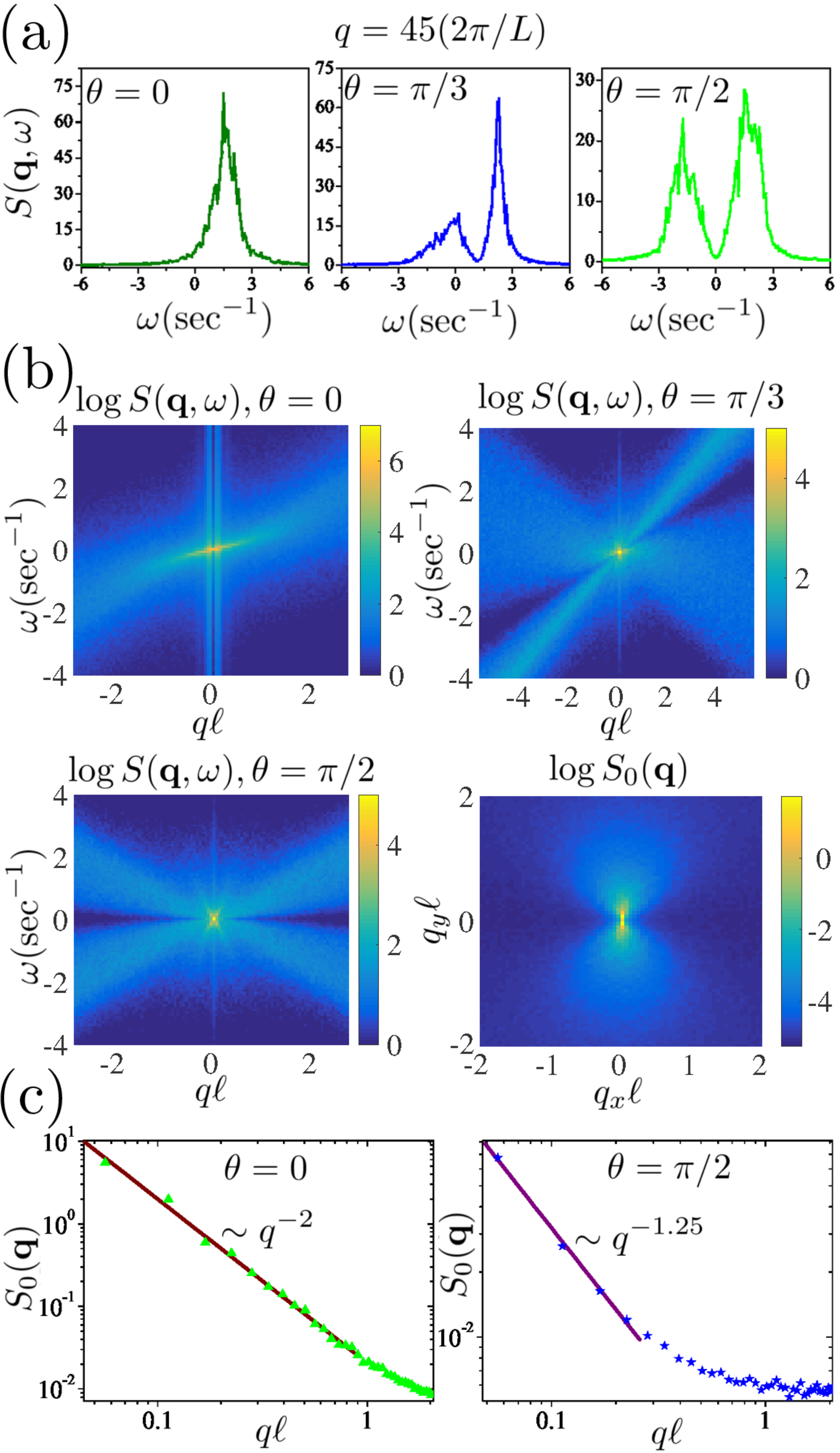} 
	\caption{\textbf{(a)} $S(\textbf{q},\omega)$ vs $\omega$ for  $q=45(2\pi/L)$ at angles $\theta=0$, $\theta=\pi/3$ and $ \theta= \pi/2$ with the direction of the flock. \textbf{(b)} Heat map of  $\log S(\textbf{q},\omega)$ as function of $q$ and $\omega$ at the three different angles same as in (a), and heat map of the static structure factor $\log S_0(\textbf{q})$ as the function of  $q_x$ and $q_y$. \hrs{\textbf{(c)} $S_0(\textbf{q})$ vs $q$ for $\theta=0$ and $\theta=\pi/2$: $S_0(\textbf{q})$ goes as $q^{-\alpha}$ with $\alpha=2$ and $\alpha=1.25$ for $\theta=0$ and $\theta=\pi/2$ respectively.} Here, the flock is moving along the $x$ direction. Parameter values: $\Phi_r=0.11$, $\Phi_b=0.60$ and $L=112$.}
	\label{wave1}
\end{figure}
where sum is taken over all the rods. The static orientation structure factor $S_0(\textbf{q})$ is defined as 
\begin{eqnarray}
S_0(\textbf{q})=\frac{1}{N} \langle \delta \textbf{p}_{\perp}(\textbf{q},t)\cdot \delta \textbf{p}_{\perp}(-\textbf{q}, t)\rangle_t,
\end{eqnarray} 
where $\langle \rangle_t$ stands for average over time $t$ and 
\begin{eqnarray}\label{09092014}
\delta \textbf{p}_{\perp}(\textbf{q}, t)=\sum_i \textbf{n}^{\perp}_{i}(t) \exp \left( -i \textbf{q} \cdot \textbf{R}^r_i(t) \right) 
\end{eqnarray}
is the Fourier  transform of $\delta \textbf{p}_{\perp}(\textbf{r}, t)$ in space. The expression for the dynamic  orientation structure factor $S(\textbf{q},\omega)$ reads
\begin{eqnarray}
S(\textbf{q},\omega)=\frac{1}{N} \langle \delta \textbf{p}_{\perp}(\textbf{q}, \omega)\cdot \delta \textbf{p}_{\perp}(-\textbf{q}, -\omega)\rangle,
\end{eqnarray}
where angle bracket stands for an average over time or configuration and $\delta \textbf{p}_{\perp}(\textbf{q}, \omega)$ is the Fourier transform of  $\delta \textbf{p}_{\perp}(\textbf{q},t)$ in time:
\begin{eqnarray}
\delta \textbf{p}_{\perp}(\textbf{q}, \omega)=\int_t dt \exp (i \omega t)\sum_i \textbf{n}^{\perp}_{i}(t) \exp \left( -i \textbf{q} \cdot \textbf{R}^r_i(t) \right).
\end{eqnarray}
\begin{figure}[htb]
	\centering \includegraphics[width=0.5\textwidth]{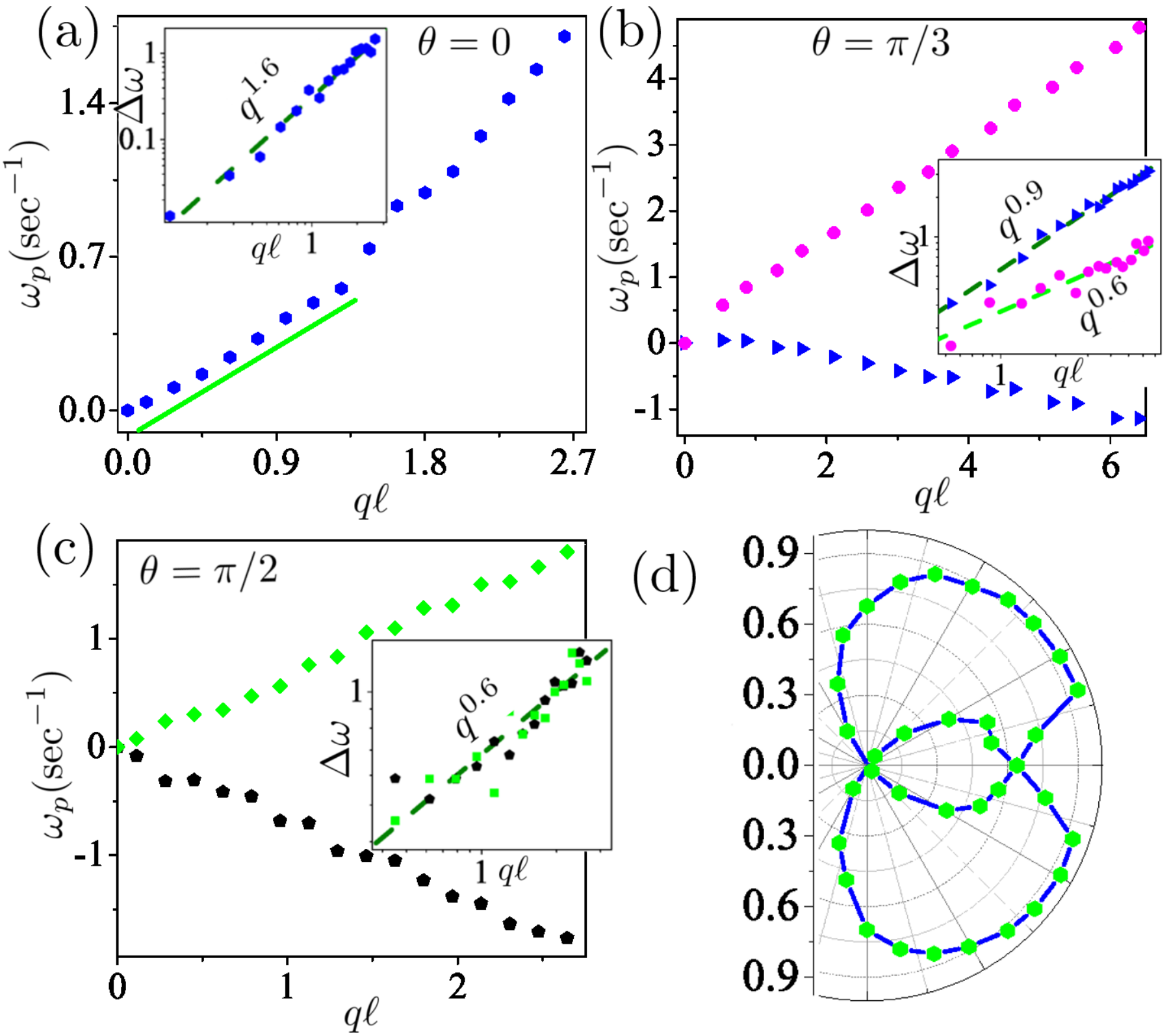} 
	\caption{ \textbf{(a)}, \textbf{(b)} and \textbf{(c)} show $\omega_p(\textbf{q})$ vs $q$ for wavevectors  $\textbf{q}$ parallel and antiparallel to the directions $\theta=0$, $\pi/3$ and $\pi/2$ with respect to the flocking direction. The insets show peak widths $\Delta \omega(\textbf{q}) \propto q^{\beta}$, with $\beta=1.6$ for $\theta=0$, and $\beta=0.6$ and $0.9$ for $\pi/3$, and $\beta=0.6$ for $\pi/2$. \textbf{(d)} A polar plot presenting the measured speed of the sound modes as a function of angle $\theta$ of propagation vector with direction of flock, calculated at $q=45(2\pi/L)$. Considerable dispersion is seen for $\textbf{q}$ antiparallel to $\pi/3$ as the speed at $q=0$ is small. Here $\Phi_r=0.11$, $\Phi_b=0.60$ and $L=112$.}
	\label{wavenew}
\end{figure}
In Fig. \ref{wave1}a,  $S(\textbf{q},\omega)$ vs $\omega$ is reported for modes with wavevectors $\textbf{q}$ parallel and antiparallel to the directions $\theta=0$, $\pi/3$ and $\pi/2$ with respect to the flocking direction, for $\Phi_r=0.11$, $\Phi_b=0.60$ and $L=112$. We also construct  the heat maps  for $S(\textbf{q},\omega)$ as the function $q$ and $\omega$ these three directions (see Fig. \ref{wave1}b): single mode for $\theta=0$ and two modes for the other cases are clearly visible. Correspondingly, $S(\textbf{q},\omega)$ as the function of $\omega$ shows two peaks in all the directions which merge into one along the direction of the flock. The heat map for the  static structure factor for orientation $S_0(\textbf{q})$ is presented in the bottom-right panel of the Fig. \ref{wave1}b, which reveals that the waves are highly anisotropic. 
\hrs{Fig.~\ref{wave1}c points out that $S_0(\textbf{q})$ goes as $q^{-\alpha}$ with $\alpha=\alpha_{\parallel}=2$ and $\alpha=\alpha_{\perp}=1.25$ along and normal to the flock, respectively. These exponent values are consistent with the Toner-Tu theory~\cite{tonertu_pre_1998} which predicts $\alpha_{\parallel}=2$ and $\alpha_{\perp}=1.2$  .}
  The speed of the propagating mode of wave vector $\textbf{q}$ is defined as
\begin{equation}
v_s(\textbf{q})=\omega_p(\textbf{q})/q,
\end{equation}
 where $\omega_p(\textbf{q})$ is the position of the peak corresponding to the mode in $S(\textbf{q},\omega)$ vs $\omega$ plot. \hrs{We find that $\omega_p$ is proportional to $q$ for small $q$ (see Fig. \ref{wavenew}a, b and c for $\theta=0$, $\pi/3$ and $\pi/2$ respectively), except where $v_s(q=0)$ is very small, e.g., wavevectors at $4\pi/3$ to the flocking direction. The width of the peak $\Delta \omega(\textbf{q})$ displays a dependence consistent with a power law $q^{\beta}$. The value of $\beta$ is $\beta_{\parallel}=1.6$ for $\theta=0$, 0.6 and 0.9 for the two peaks at $\theta=\pi/3$ and $\beta_{\perp}=0.6$ for both the peaks at $\theta=\pi/2$ (insets of Fig.~\ref{wavenew}a-c).  This is in contrast to the Toner-Tu~\cite{tonertu_pre_1998} prediction $\beta_{\parallel}=2$ and $\beta_{\perp}=1.2$.}  
  Fig. \ref{wavenew}d shows the angular dependence of $v_s(\theta)$ at $q=45 (2 \pi/ L)$. Two loops intersecting at $\theta=0$ correspond to two wave modes. All these observations fairly agree with the theoretical predictions~\cite{tonertu_pre_1998} except the wavenumber dependence of the sound peak widths. These waves have been already detected in the numerical simulations of Vicsek-style models~\cite{soundwave_prl_1997} but observing them in our numerical model, which is mechanically realistic rather than agent-based, suggests that these modes should exist in real granular-matter experiments. For a systematic evaluation of the Toner-Tu predictions against large-scale numerical studies of the Vicsek model see \cite{PhysRevLett.123.218001}.

\subsubsection{Large density fluctuations}~\label{largN}
\begin{figure}[H] 
	\centering \includegraphics[width=0.5\textwidth]{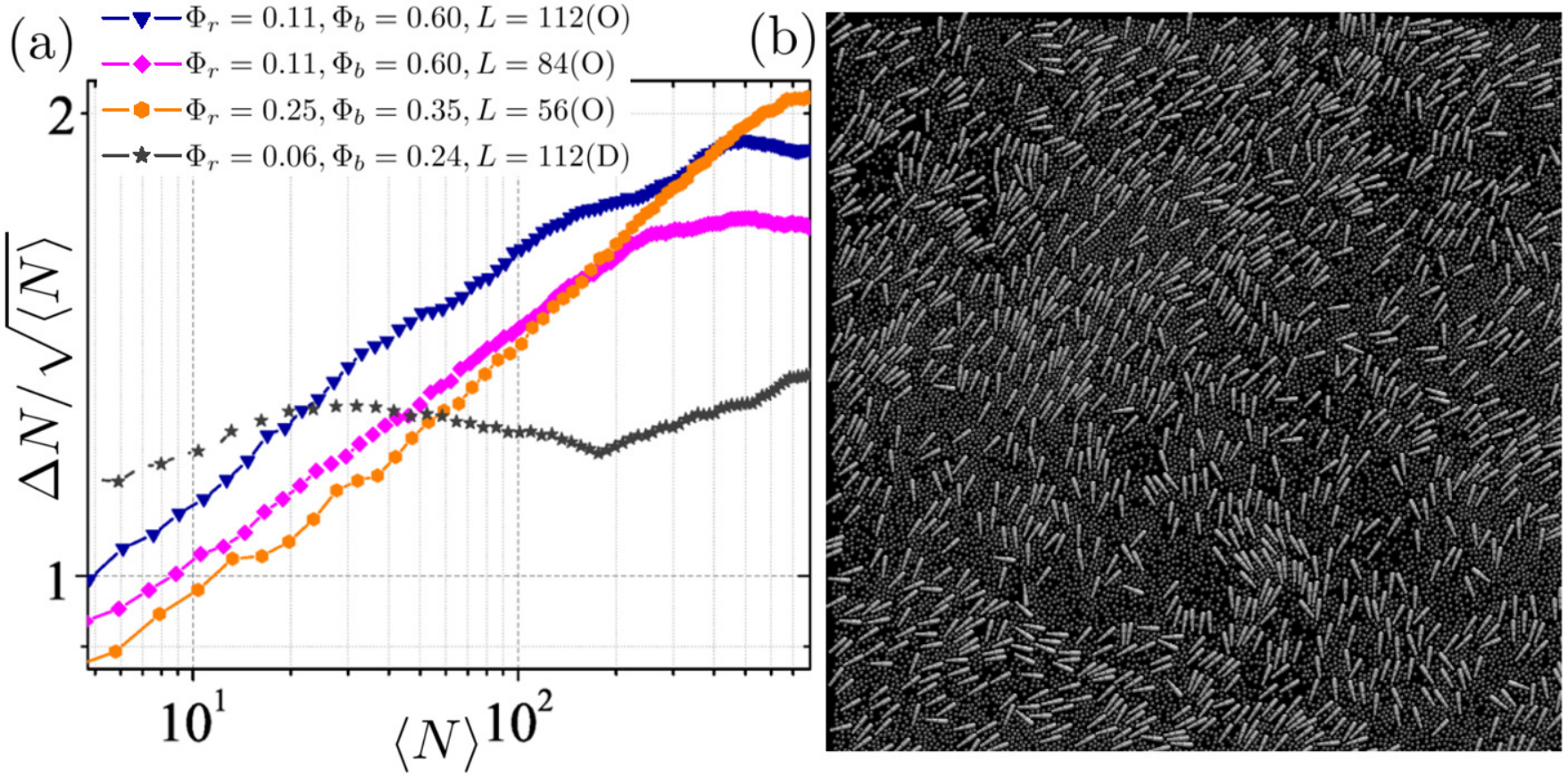} 
	\caption{\textbf{(a)}. Variance in number of rods scaled by $\sqrt{\left\langle N\right\rangle }$ vs average number of rods $\left\langle N\right\rangle $ for three ordered states (labelled O) and one disordered state (labelled D). \textbf{(b)} The snapshot of the ordered state at $\Phi_r=0.25$ and $\Phi_b=0.35$ at $ L=56$.}
	\label{nufluct}
\end{figure} 
In the thermal equilibrium systems away from critical points, in the thermodynamic limit, the particle number fluctuations in the grand canonical ensemble grow as the square root of the average number of the particle $\langle N\rangle$ as the system size is increased \textit{ i.e.} root-mean-square deviation 
\begin{eqnarray}
\Delta N =\sqrt{\langle N^2 \rangle-\langle N \rangle^2} \propto \langle N \rangle^{1/2},
\end{eqnarray}
where $N$ is the instantaneous number of the particles and $\langle  \rangle$ represents the ensemble average for a given system size. In contrast, active systems show anomalous number fluctuation properties: hydrodynamical theories~\cite{tonertu_pre_1998,sriram_aditi_2002,sr_epl_2003} suggest that  $\Delta N$ for active polar and apolar systems is proportional to $\langle N \rangle^{1/2}$ in isotropic states but the broken-symmetry states demonstrate large number fluctuations growing as ~\cite{tonertu_pre_1998,sriram_aditi_2002,sr_epl_2003}
\begin{eqnarray}\label{numfluc1}
\Delta N \propto \langle N \rangle^{\eta+\frac{1}{2}},
\end{eqnarray} 
where $\eta= 1/d$ in mean field theory and has smaller values in 
renormalization-group treatments of active polar~\cite{tonertu_pre_1998} and nematic~\cite{PhysRevE.97.012707} phases.
We construct series of the number of the particles inside the subsystems of different sizes co-centred with the original system, from many statistically independent realizations. We then calculate $\langle N \rangle$ and $\Delta N$ for each subsystems.
At low $\Phi_r$, in the disordered isotropic state, $\Delta N$ is proportional to $\langle N \rangle^{1/2}$ (see Fig. \ref{nufluct}a for $\Phi_r=0.06$ and $\Phi_b=0.24$). Fig. \ref{nufluct}a presents $\Delta N/\sqrt{\langle N \rangle}$ vs $\langle N \rangle$ for three different ordered states: the number fluctuations are larger than the thermal systems  with $a \sim 0.6$ but weaker than the ones predicted by Toner-Tu theory. More detailed studies with finite-size scaling are required for a definitive exponent estimate. Here we should make it clear that the large number fluctuations in order states are not due to a segregation as seen in Fig.~\ref{phasediagram}e \& h: a typical ordered state has homogenous rod density (see  Fig. \ref{nufluct}b).

\begin{figure}
	\centering \includegraphics[width=0.3\textwidth]{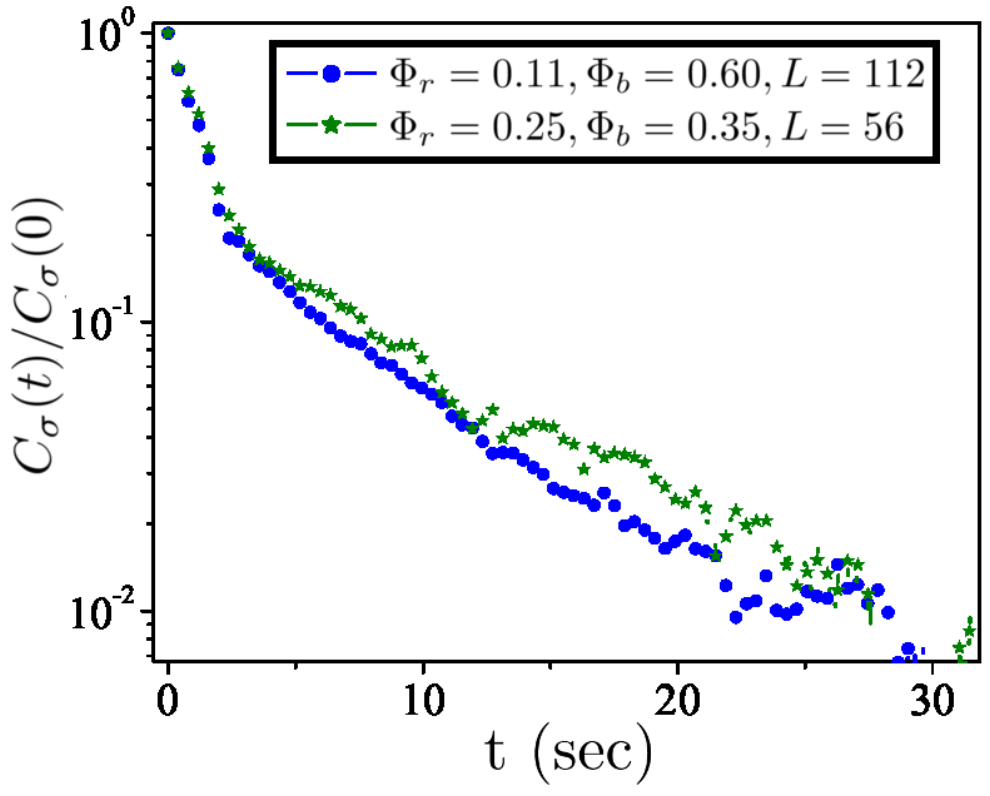} 
	\caption{\hrs{The density autocorrelation function $C_{\sigma}(t)$ scaled by its value at $t=0$ as the function of $t$.  Here $\Phi_r=0.11$, $\Phi_b=0.60$ and $L=112$ and the size of the cell used for coarse-graining is 2$\ell$.}}
	\label{denaut}
\end{figure}

In order to quantify the lifetime of the number fluctuations we also calculate the density autocorrelation function defined as
\begin{equation}
C_{\sigma} (t)= \left\langle (\sigma (\textbf{r},t) -\sigma_0(\textbf{r})) (\sigma (\textbf{r},0) -\sigma_0(\textbf{r}))\right\rangle, 
\end{equation}
where $\sigma_0(\textbf{r})$ is the average density at the position $\textbf{r}$  and $\left\langle \right\rangle$ represent the average over space. \hrs{The coarse-grained number density for the rods $\sigma (\textbf{r},t)$ has been defined in~\ref{band_section}.Fig. \ref{denaut} suggests that $C_{\sigma}(t)$ shows two exponential decays. We do not understand this observation. A linearized treatment as in Narayan \textit{et al.}~\cite{Narayan06072007} would give a logarithmic decay, while the Toner-Tu theory~\cite{tonertu_pre_1998} would give a slower $t^{-2/3}$ in two dimensions. Neither of these forms gives a reasonable fit to our data.}

\subsection{Experiments}~\label{exp}
\begin{figure*}[t]
	\centering \includegraphics[width=1.0\textwidth]{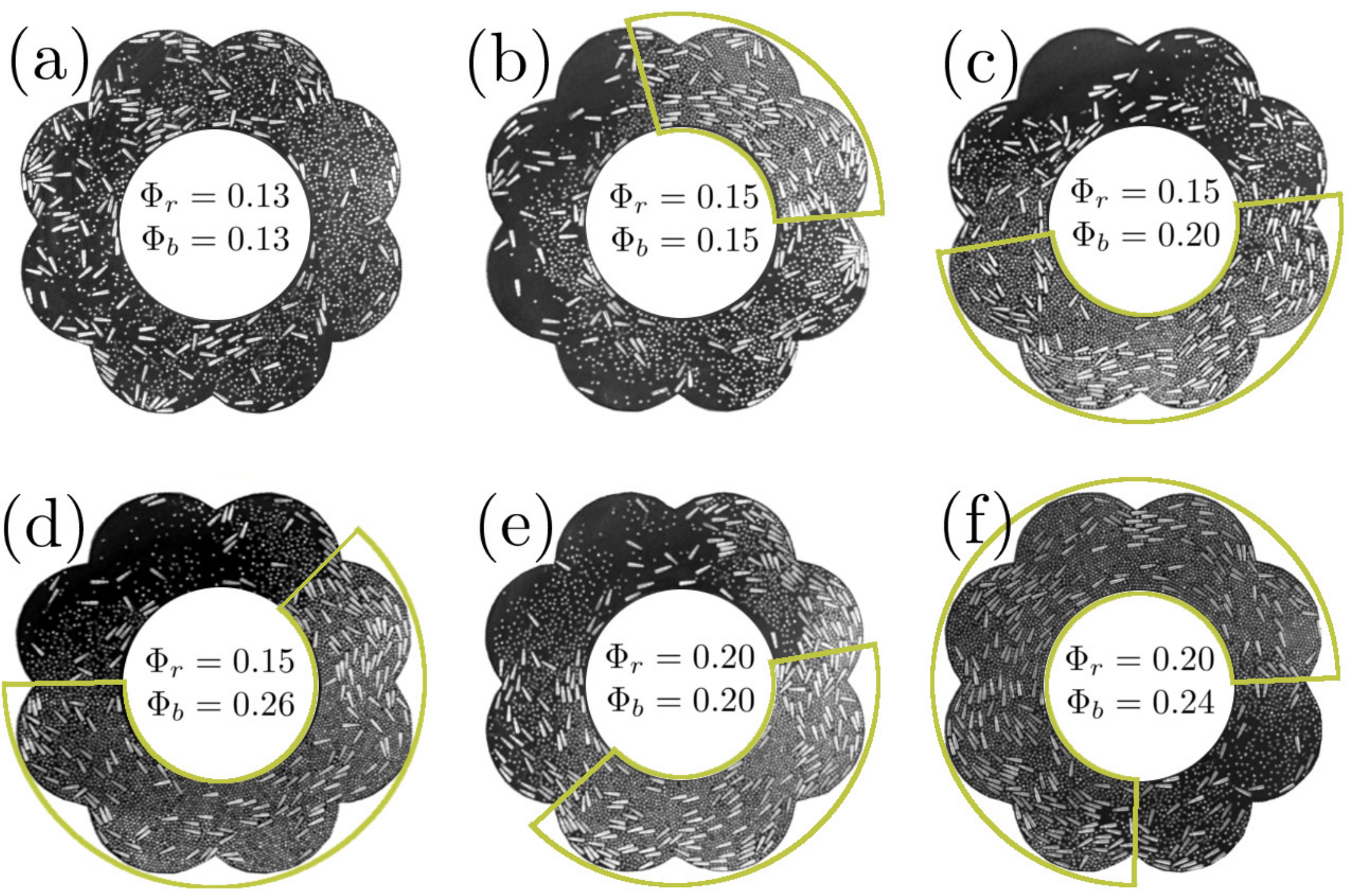} 
	\caption{Images from the experiments performed with rods and beads in annular geometry: (a) an isotropic phase and (b-d) bands at different $\Phi_r$ and $\Phi_b$. It can be seen from (b-d) for $\Phi_r=0.15$, and (e-f) for $\Phi_r=0.20$, that the size of the band increases with $\Phi_b$, as observed in our simulations. }
	\label{expxpband}
\end{figure*}
\hrs{As we showed in earlier work~\cite{kumar2014flocking}, a flocking transition can be triggered at very low area fraction of motile polar rods by increasing the concentration of non-motile beads (see Fig.~\ref{exp} of Appendix). Here, motivated by our simulations, we perform the experiments in annular geometry at varying $\Phi_b$ and $\Phi_r$ and find that for parameter values shortly past onset of flocking a band does emerge, with appearance typically as in Fig.~\ref{expxpband}b-f and Supplementary Movie 5. The band is identified as the area of high density of rods and beads with high degree of orientational order which is moving in the direction of its order, with a background of isotropic mixture with low concentration of both rods and beads.
Fig.~\ref{expxpband} indicates that, at fixed $\Phi_r$, the band becomes wider as $\Phi_b$ increased, which is consistent with our simulation results. The location of the bands in the $\Phi_r-\Phi_b$ plane is depicted in the phase diagram in Fig. \ref{phasediagram}a. Although the observation of band formation in our experiments is robust, limitations on system size make a more systematic study impractical.}
\section{Future directions and summary}\label{summary}
The research presented in this article was motivated by multiple aims. One was to study collective motion with both self-propulsion and alignment arising for purely mechanical rather than behavioural reasons. The second was to carry out such a study in a system in which -- unlike in the Quinke rollers \cite{Bricard2013} -- the strength of interactions could be tuned simply, in this case by changing the bead concentration. Third was to see if the band formation just past ordering onset could be modulated or suppressed. Despite the limitations on system size, we believe we have succeeded in the above aims. Two regimes remain to be explored in more detail. One is the limit of no beads, \SR{not discussed here} where a patchy MIPS precursor possibly leads to a dilute flock at low noise (see Fig.~\ref{snaps_nobeads} of Appendix and Supplementary Movie 6). The other is the bead-rod phase separation that occurs at high area fraction of both species.

We close with a summary. In our detailed numerical simulations, we observed four distinct phases of our system: disordered, banded and ordered, homogeneous and ordered, and bead-rod phase separated. The existence of the broken symmetry state at very low rod concentration is a key result. The banded state is similar to the one seen in a system of the polar disks~\cite{PhysRevLett.110.208001} or in the rolling colloids \cite{Bricard2013}: only single bands are observed which are generally aligned with the boundary of the periodic box but also could be in an arbitrary direction at high rod concentration.  This is in contrast to the density waves exhibited by the Vicsek  model~\cite{chate_bands_2014}. Moreover, in these systems, the band absorbs the bead medium as well, for which there is no equivalent in the other systems like polar disks and colloidal rollers~\cite{Bricard2013}. The width of the bands increases with the rod concentration. 
We also find the first experimental realisation of bands in a dry granular flocking system. We also explore the anisotropic propagating waves and large number fluctuations in the highly ordered state. Some interesting departures are observed with respect to the Toner-Tu theory~\cite{tonertu_pre_1998}, especially as regards the scaling of damping rate with wavenumber. 

\section*{Conflicts of interest}
There are no conflicts to declare.
\FloatBarrier
\section*{Appendix}

\begin{figure}[H] 
	\centering \includegraphics[width=0.5\textwidth]{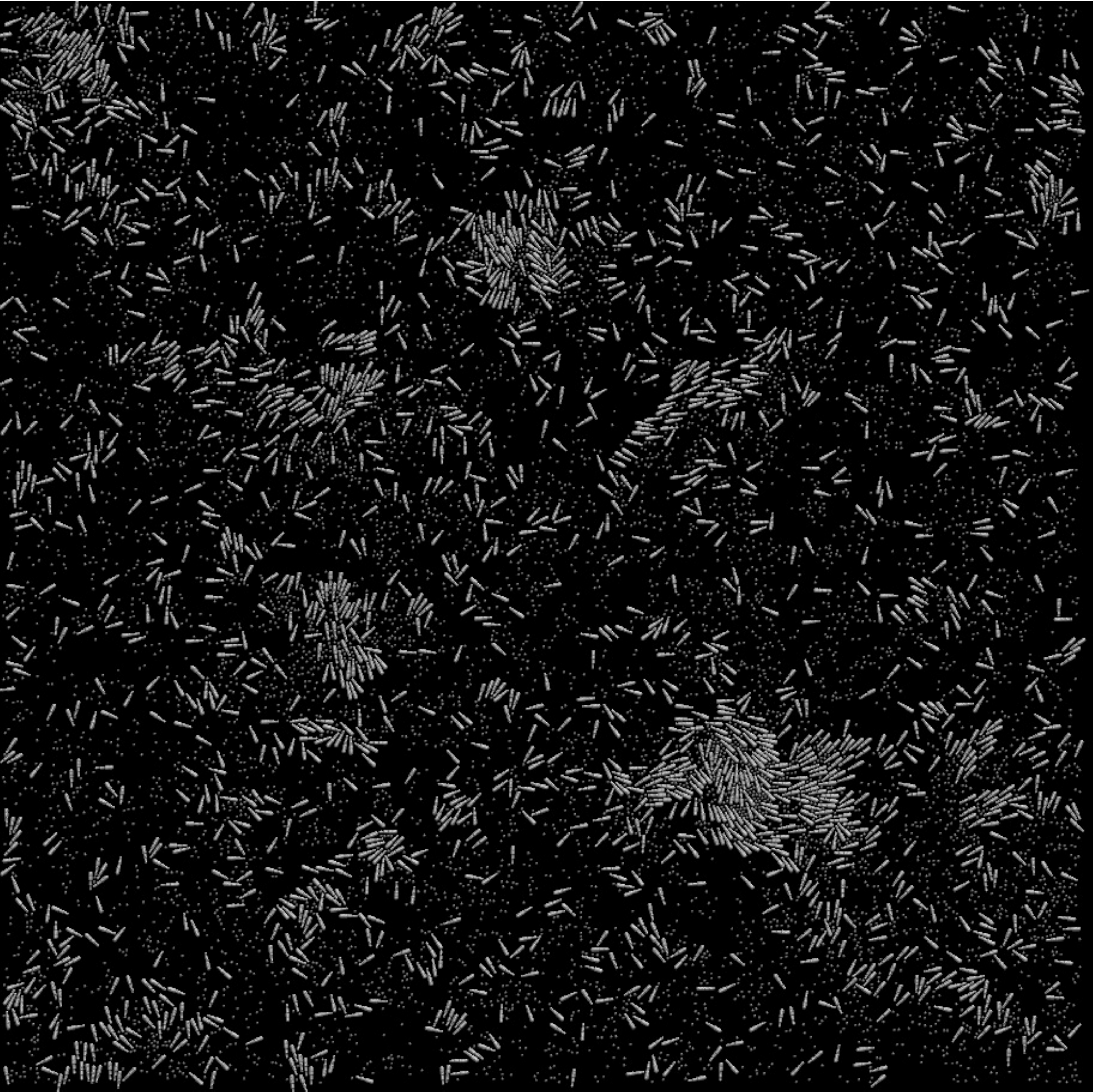}
	\caption{A disordered phase at low $\Phi_b$ and high $\Phi_r$ having locally ordered swarms. Here, $\Phi_r=0.13$ and $\Phi_b=0.07$.}
	\label{disorder13_07}
\end{figure}
\begin{figure}[H] 
	\centering \includegraphics[width=0.3\textwidth]{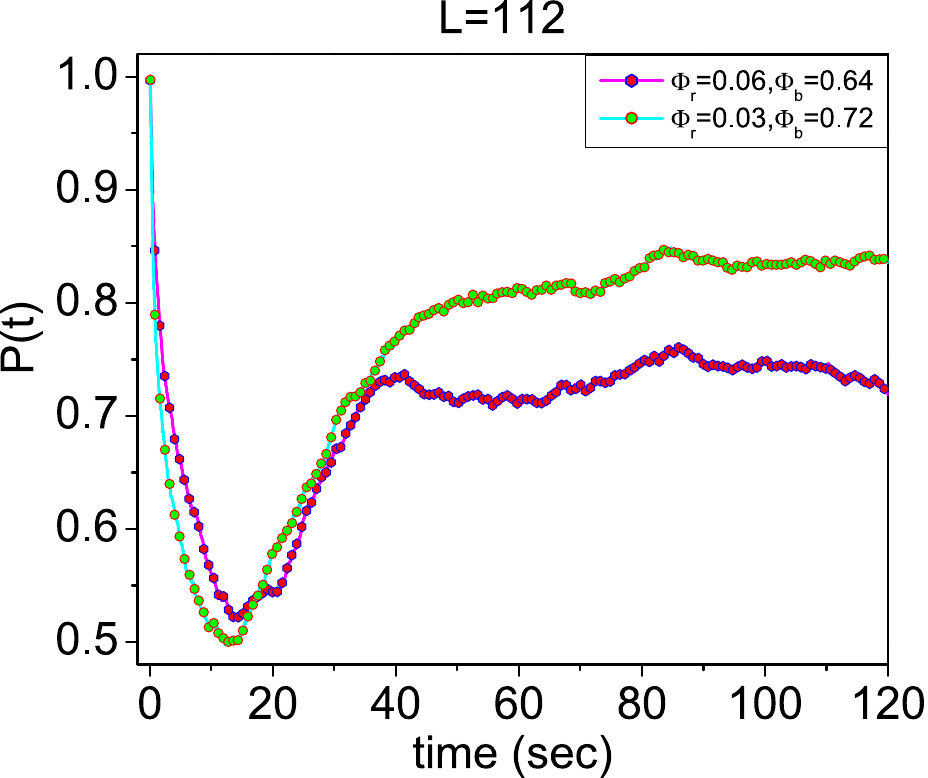}
	\caption{The time evolution of the polar order parameter $P(t)$ of the system which is perfectly ordered with zero velocity at $t=0$. Initially, the ordering of the rods starts lessening due to the absence of the bead flow. In some time, as soon as, the beads picks up the velocity as they are pushed and dragged by the rods, the rods begin to realign due to the interaction with the bead flow. Correspondingly the order parameter $P(t)$ initially decreases and again increases after some time.}
	\label{rolebeads}
\end{figure}
\begin{figure}[H] 
	\centering \includegraphics[width=0.5\textwidth]{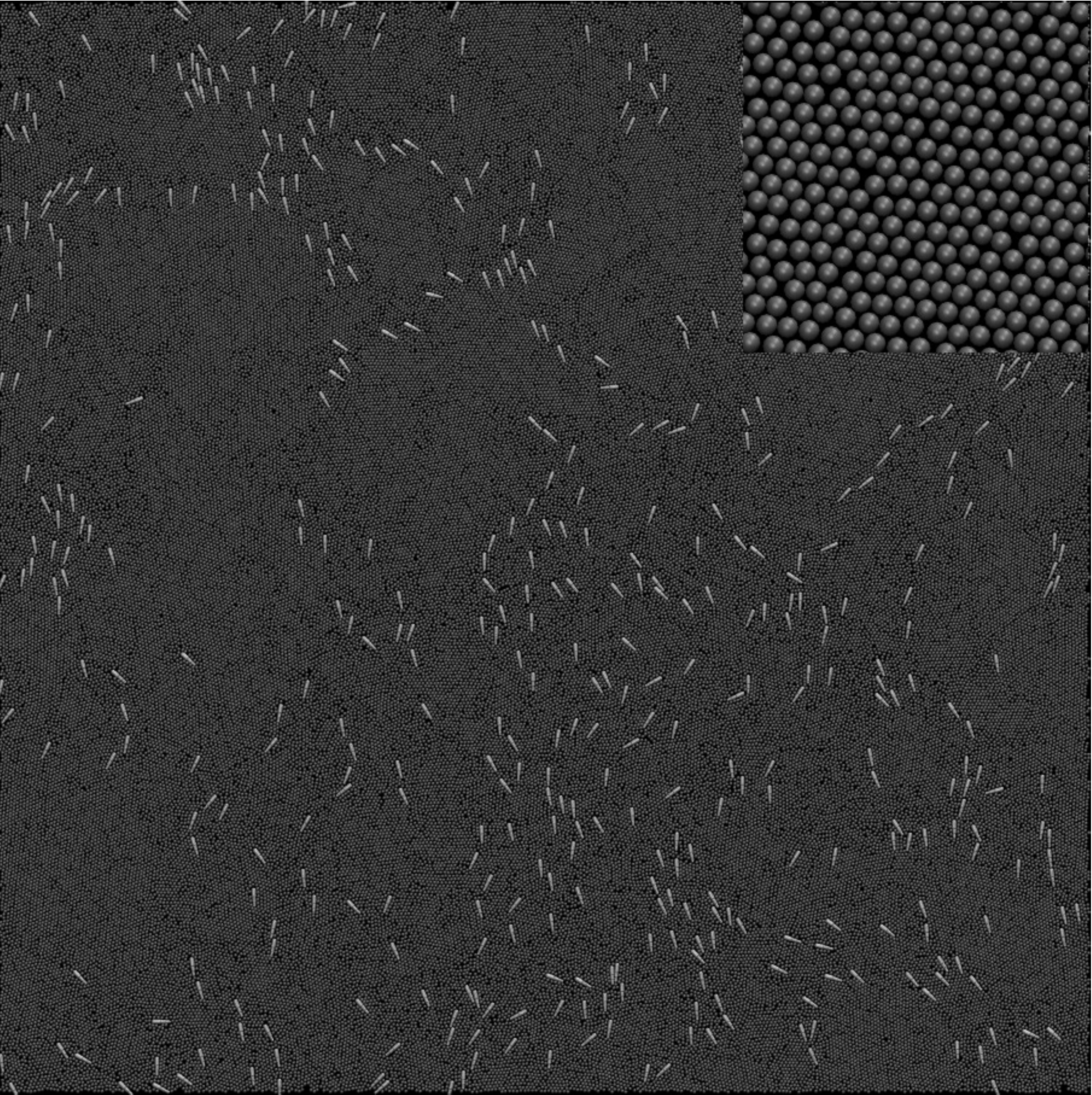}
	\caption{The ordered state at extremely low rod density $\phi_r= 0.03$ with
		$\phi_b = 0.72$. The inset shows the crystalline order in the bead medium.}
	\label{r03t75}
\end{figure}
\begin{figure}[H]
	\centering \includegraphics[width=0.5\textwidth]{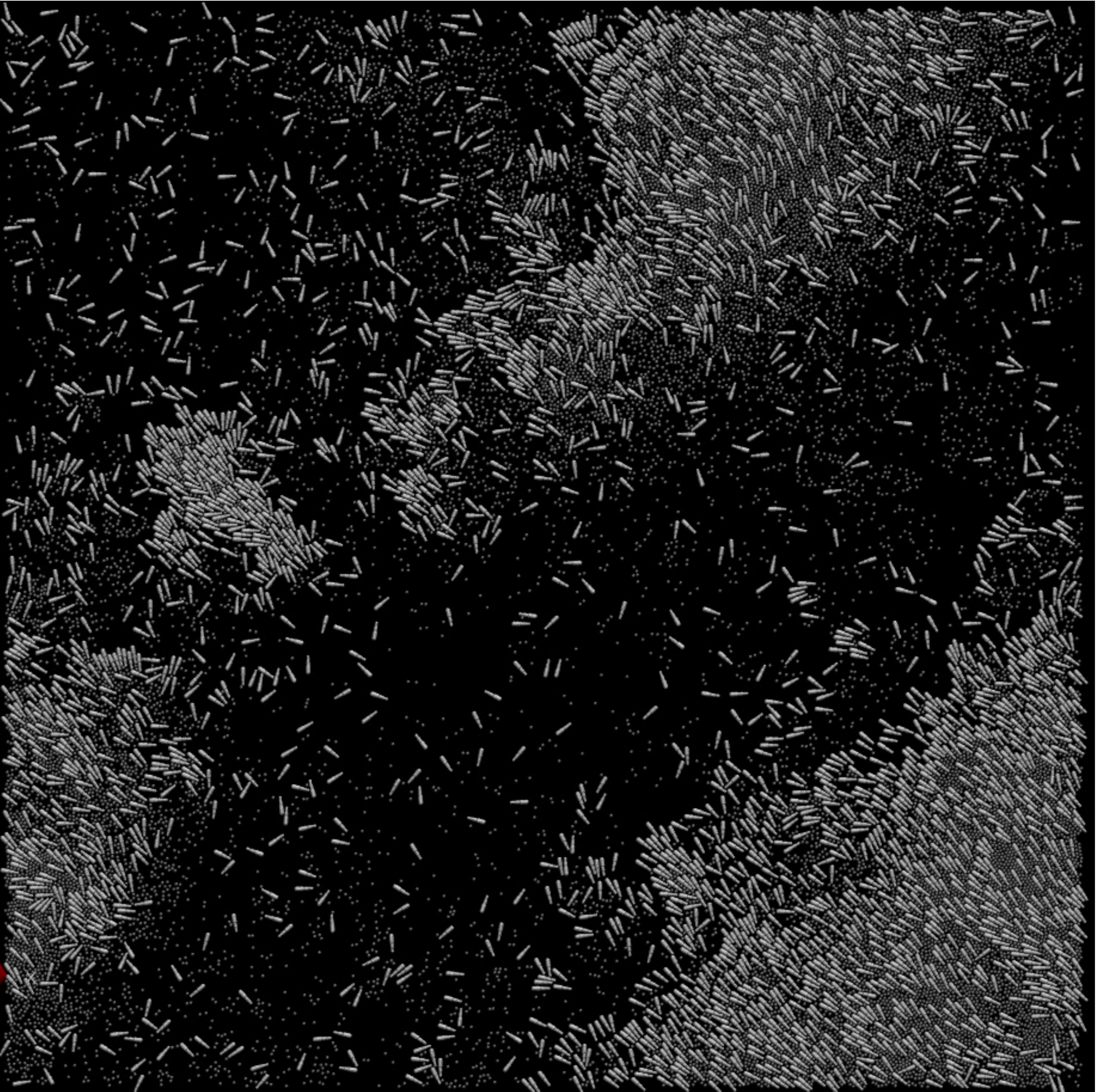} 
	\caption{A band moving in a direction not parallel to the sides of the periodic box at $\phi_r=0.19$ and $\Phi_b=0.11$. }
	\label{diagonalbands}
\end{figure}
\begin{figure}[H] 
	\centering \includegraphics[width=0.5\textwidth]{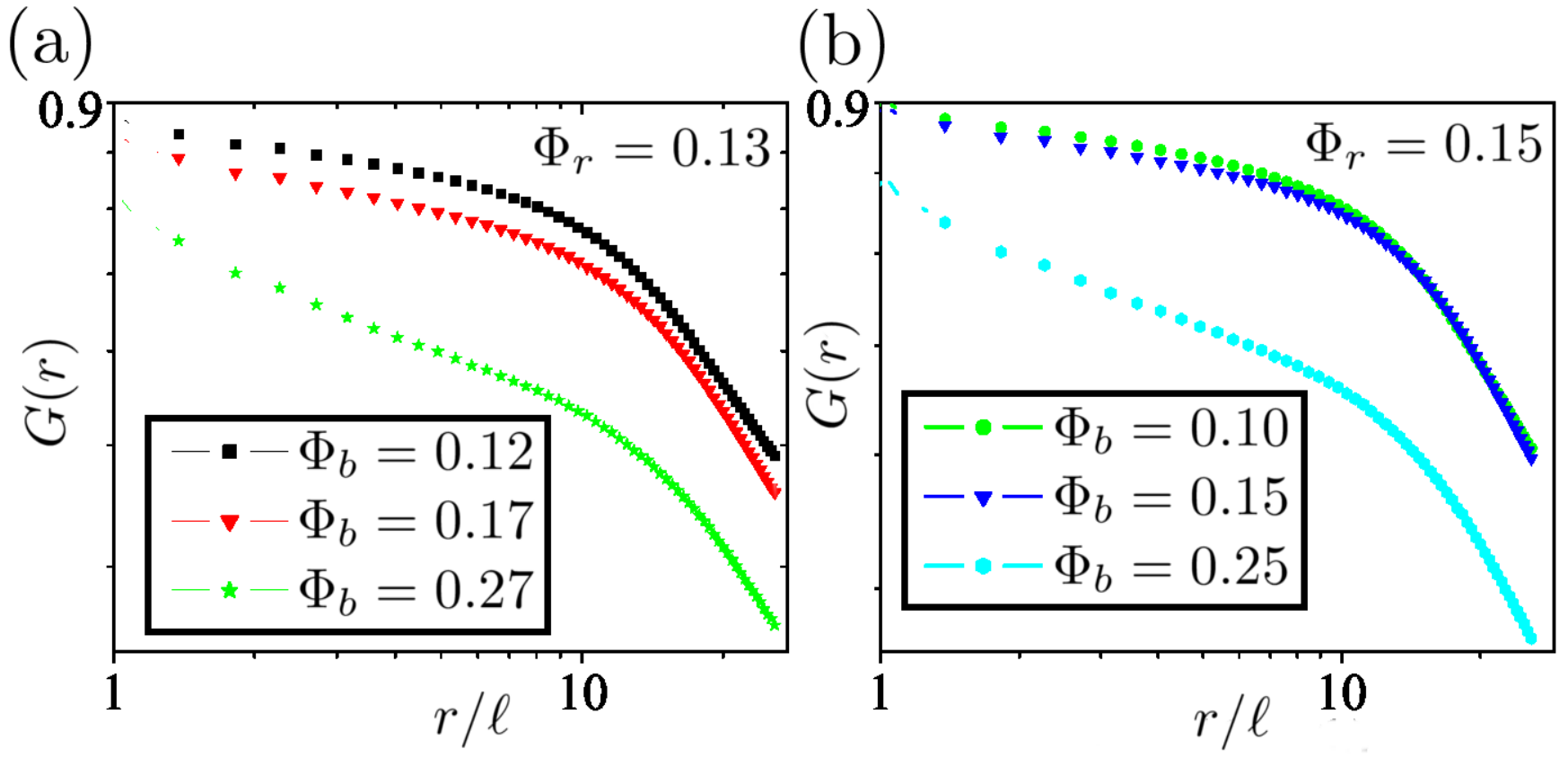} 
	\caption{Order parameter correlation function $G(r)$ vs $r$ in the band regime of the Phase diagram.  $\Phi_r=0.13$ for \textbf{(a)} and  $\Phi_r=0.15$ for \textbf{(b)}. System size $L=56$.}
	\label{bandallgr}
\end{figure}
\begin{figure}[H] 
	\centering \includegraphics[width=0.5\textwidth]{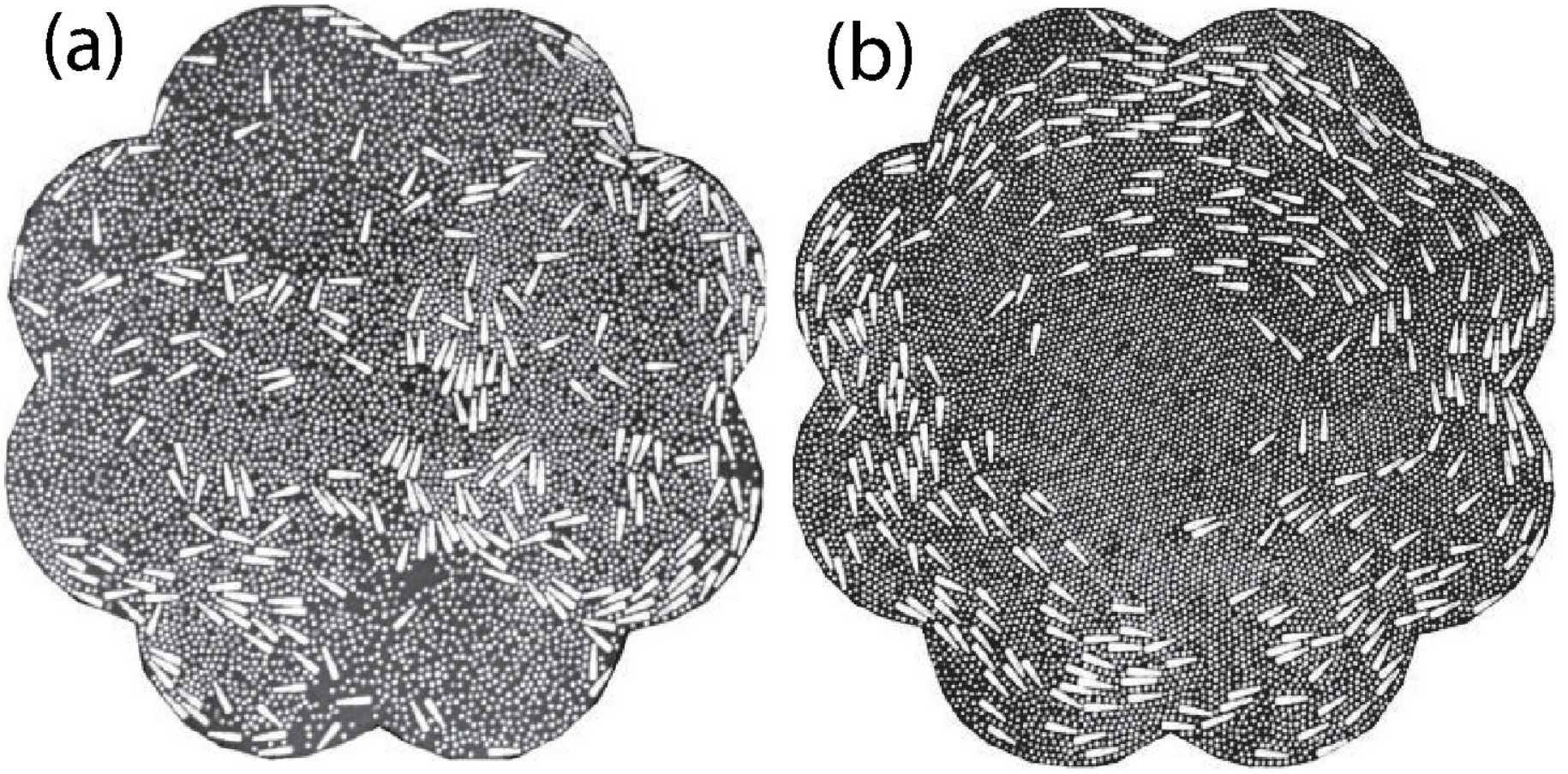}
	\caption{ In experiments, a disordered phase at low $\Phi_b$ (left), and an ordered phase at high $\Phi_b$ (right) of a system of rods and beads~\cite{kumar2014flocking}.}
	\label{exp}
\end{figure}
\begin{figure}[H] 
	\centering \includegraphics[width=0.5\textwidth]{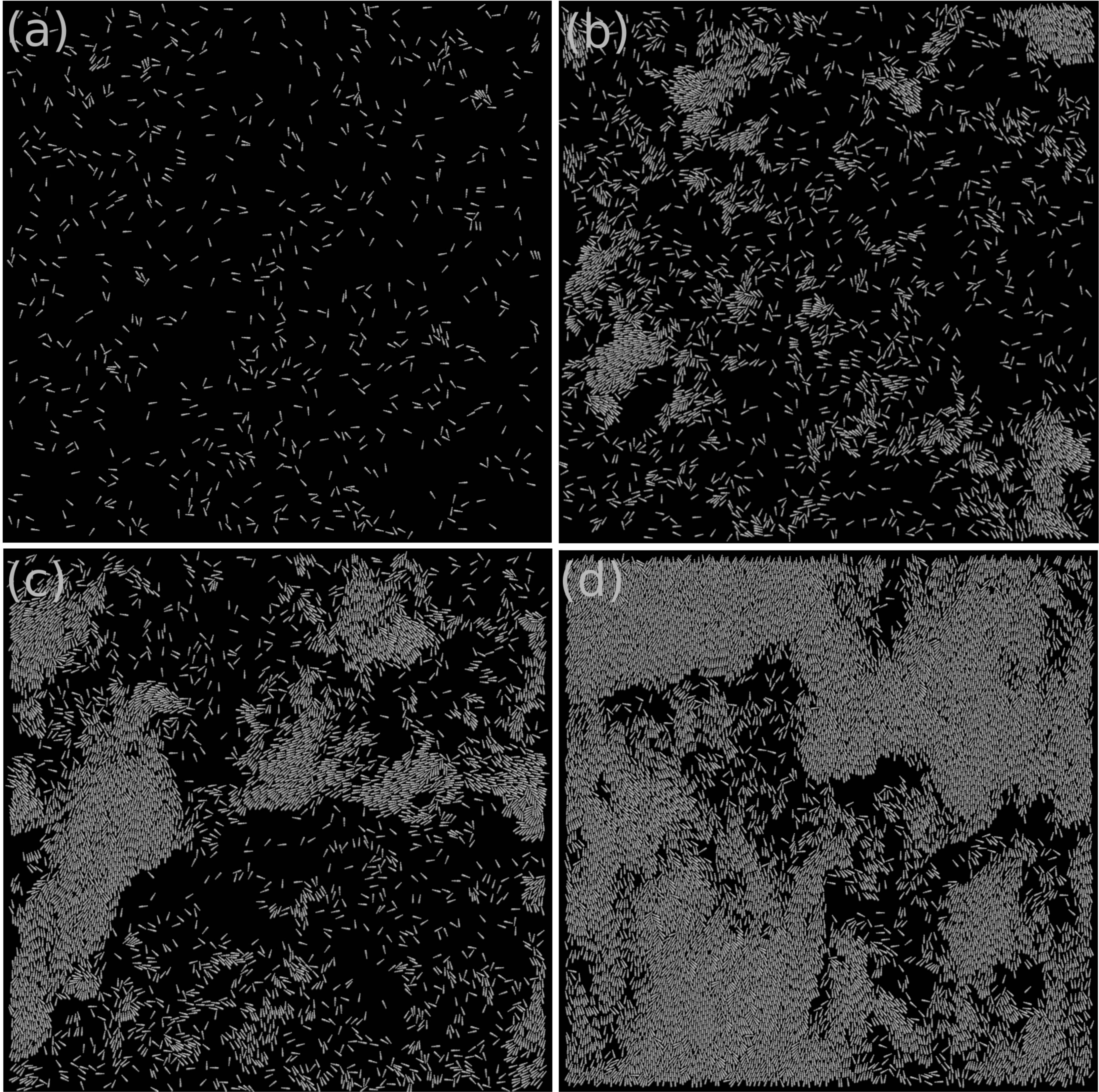} 
	\caption{ Behaviour of the system with no beads. \textbf{(a)} Disordered state at $\Phi_r=0.05$. Patchy swarms  at large $\Phi_r$:  \textbf{(b)} at $\Phi_r=0.20$,  \textbf{(c)} at $\Phi_r=0.30$ and \textbf{(d)} at $\Phi_r=0.60$. Size of the swarms increases with $\Phi_r$.}
	\label{snaps_nobeads}
\end{figure} 

\section*{Acknowledgements}
NK, HS and JN thank the UGC, the CSIR and the Science Academies' Summer Research Fellowship respectively for support. AKS was supported by a Year of Science Professorship by Department of Science and Technology, India and SR by the Tata Education and Development Trust and a J C Bose Fellowship of the SERB. 
 
\balance

\FloatBarrier
\providecommand*{\mcitethebibliography}{\thebibliography}
\csname @ifundefined\endcsname{endmcitethebibliography}
{\let\endmcitethebibliography\endthebibliography}{}

\bibliographystyle{rsc} 

\begin{mcitethebibliography}{58}
\bibitem[sep()]{separationfoot}
	In the present context MIPS is simply joint condensation of rods and beads, and
	should not be confused with the demixing of the two species, which we call
	segregation, observed at higher area fraction\relax
	\mciteBstWouldAddEndPuncttrue
	\mciteSetBstMidEndSepPunct{\mcitedefaultmidpunct}
	{\mcitedefaultendpunct}{\mcitedefaultseppunct}\relax
	\EndOfBibitem
\end{mcitethebibliography}

\end{document}